\documentclass[aps,superscriptaddress,twocolumn]{revtex4}
\usepackage{feynmp}
\usepackage{amsmath}
\usepackage{amssymb}
\usepackage{epsfig}
\usepackage{graphicx}
\usepackage{subfigure}
\usepackage{hyperref}
\usepackage{bbm}
\usepackage{color}
\usepackage{epstopdf}

\newcommand{\Rmnum}[1]{\expandafter\@slowromancap\romannumeral #1@}
\allowdisplaybreaks[3]

\begin{document}

\title{Phase transition with trivial quantum criticality in anisotropic Weyl semimetal}

\author{Xin Li}
\affiliation{Department of Modern Physics, University of Science and
Technology of China, Hefei, Anhui 230026, P. R. China}
\author{Jing-Rong Wang}
\affiliation{Anhui Province Key Laboratory of Condensed Matter
Physics at Extreme Conditions, High Magnetic Field Laboratory of the
Chinese Academy of Science, Hefei, Anhui 230031, P. R. China}
\author{Guo-Zhu Liu}
\altaffiliation{Corresponding author: gzliu@ustc.edu.cn}
\affiliation{Department of Modern Physics, University of Science and
Technology of China, Hefei, Anhui 230026, P. R. China}

\begin{abstract}
When a metal undergoes continuous quantum phase transition, the
correlation length diverges at the critical point and the quantum
fluctuation of order parameter behaves as a gapless bosonic mode.
Generically, the coupling of this boson to fermions induces a
variety of unusual quantum critical phenomena, such as non-Fermi
liquid behavior and various emergent symmetries. Here, we perform a
renormalization group analysis of the semimetal-superconductor
quantum criticality in a three-dimensional anisotropic Weyl
semimetal. Surprisingly, distinct from previously studied quantum
critical systems, the anomalous dimension of anisotropic Weyl
fermions flows to zero very quickly with decreasing energy, and the
quasiparticle residue takes a nonzero value. These results indicate
that, the quantum fluctuation of superconducting order parameter is
irrelevant at low energies, and a simple mean-field calculation
suffices to capture the essential physics of the superconducting
transition. We thus obtain a phase transition that exhibits trivial
quantum criticality, which is unique comparing to other invariably
nontrivial quantum critical systems. Our theoretical prediction can
be experimentally verified by measuring the fermion spectral
function and specific heat.
\end{abstract}

\maketitle


\section{Introduction}

Weakly interacting metals are perfectly described by the Fermi
liquid (FL) theory \cite{GiulianiBook, ColemanBook, Shankar94}.
Coulomb interaction plays a negligible role since it becomes
short-ranged due to the static screening caused by the collective
particle-hole excitations. The static screening factor serves as an
infrared cutoff for the transferred energy/momentum, which
suppresses forward scattering and guarantees the stability of FL
state. When gapless fermions couple to certain gapless bosonic mode,
Landau damping could be strong enough to yield a vanishing
quasiparticle residue $Z_f$, which implies the breakdown of FL
theory. A prominent example is the system of fermions coupled to a
U(1) gauge boson \cite{Holstein73, Lee89, Lee92, Gan93, Altshuler94,
Polchinski94, SSLee09}. The gauge boson is strictly gapless,
rendered by local gauge invariance, and leads to non-FL behavior
characterized by $Z_f = 0$.

When a metal undergoes a continuous quantum phase transition, non-FL
behavior and other intriguing physical properties can emerge
\cite{Varma02, Lohneysen07}. Near the quantum critical point (QCP),
the quantum fluctuation of order parameter becomes critical as the
correlation length $\xi$ diverges, and can be described by the
dynamics of gapless bosonic mode \cite{Hertz76, Millis93,
Sachdev03}. The low-energy behavior of the quantum criticality is
determined by the coupling between gapless fermionic and bosonic
degrees of freedom. Such coupling has been studied extensively in
various quantum critical systems, including ferromagnetic (FM) QCP
\cite{Rech06, Ridgway15, XuMengZiYangGroup}, antiferromagnetic (AFM)
QCP \cite{Abanov03, Metlitski10AFM, Schlief17}, and Ising-type
nematic QCP \cite{Metzner03, Metlitski10Nematic, Garst10,
Lederer17}. In these systems, the fermion-boson coupling can
generate a finite anomalous dimension for fermion field and also
leads to strong Landau damping of fermions. At finite temperature,
the QCP becomes a finite quantum critical regime, as schematically
shown in Fig.~1(a), which can be called a NFL regime due to the
strong violation of FL description. A popular notion is that, the
observed superconducting (SC) dome and NFL normal-state properties
in many cuprate, heavy fermion, and iron-based superconductors arise
from the quantum fluctuation of certain long-range order.

Nontrivial quantum criticality also occurs in several semimetal (SM)
materials \cite{Vafek14, Wehling14, Armitage17, Burkov16, Yan17,
Hasan17, Weng16, FangChen16, Kotov12}. Recently, SC transition and
the associated quantum criticality have attracted particular
research interest. In most SMs, Cooper pairing occurs only when the
net attraction is larger than certain critical value \cite{Uchoa05,
Zhao05, Uchoa07, Kopnin08, Honerkamp08, Roy10, Roy13, Nandkishore13,
WangJing17, Roy16, Meng12, Maciejko14, Wei14, Sur16, Roy17,
Boettcher16, Boettcher17, Mandal17, Uchoa17, RoyFoster17}. It is
argued that the Yukawa coupling between gapless Dirac/Weyl fermions
and bosonic SC order parameter might dynamically generate an
emergent space-time supersymmetry \cite{Lee07, Ponte14, Grover14,
Witczak-Krempa16, Zerf16, Jian15, Jian17}. These QCPs display a
series of unusual quantum critical behaviors.

In this paper, we study the quantum criticality of SM-SC transition
in a 3D anisotropic Weyl semimetal (AWSM), where the fermion
dispersion is linear in two momentum exponents and quadratical in
the third one \cite{Yang13, Yang14A, Yang14B}. Such an AWSM state
emerges naturally as one pair of Weyl points of a Weyl SM merge into
one single band-touching point. In the parent Weyl SM, the Chern
numbers of one pair of Weyl points are $\pm 1$. When two Weyl points
with opposite Chern numbers merge, the resultant band-touching point
has zero Chern number \cite{Yang13, Yang14A, Yang14B}. Thus, the
AWSM state is topologically trivial. Superconductivity is induced
when the strength of net attraction, denoted by $g$, is larger than
$g_c$. At $g = g_c$, the quantum fluctuation of SC order parameter
is gapless and couples to gapless Weyl fermions. According to
previous research experience, one would naively expect to observe a
series of unusual quantum critical phenomena at the QCP.

\begin{figure}[htbp]
\center
\includegraphics[width=2.6in]{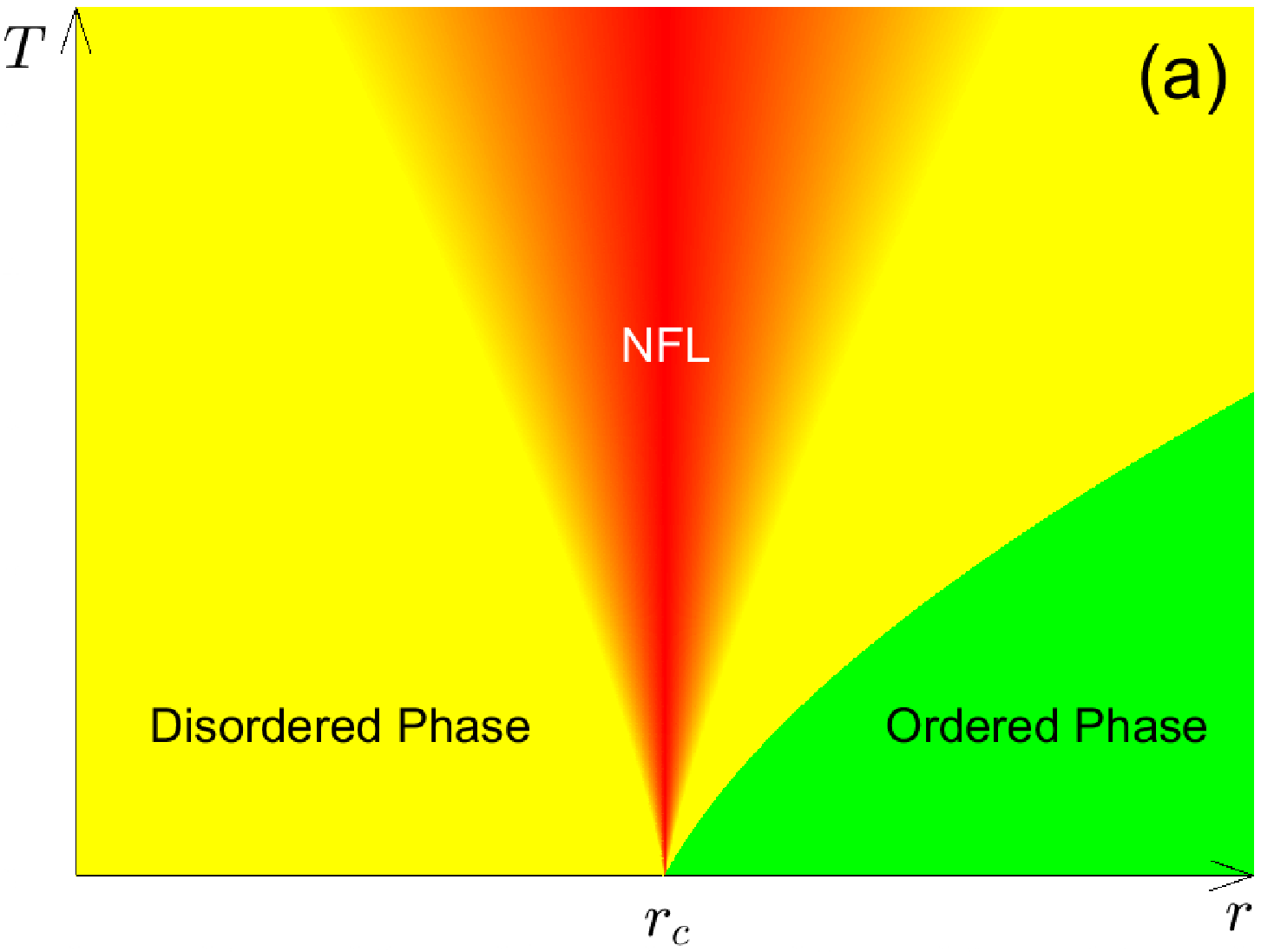}
\includegraphics[width=2.6in]{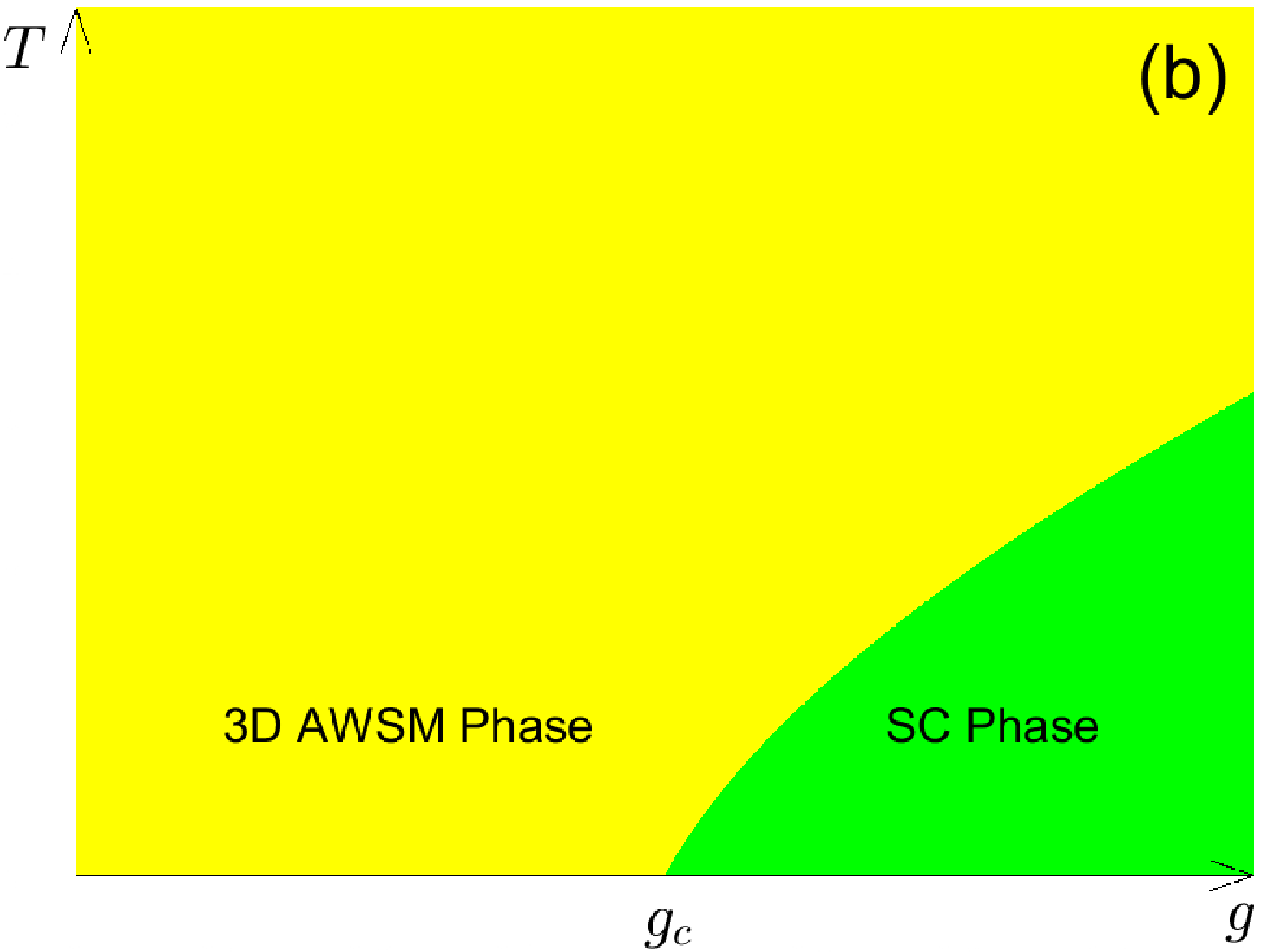}
\caption{(a) Conventional quantum critical systems always have a
large area of NFL region on the phase diagram. Here, $r$ is a tuning
parameter. (b) SM-SC QCP in 3D AWSM is trivial, because the system
exhibits qualitatively the same low-energy behavior in the whole
non-SC phase. \label{Fig:Phase}}
\end{figure}

We present a renormalization group (RG) study of the coupling
between the Weyl fermions and the SC quantum fluctuation.
Interestingly, although the SC quantum fluctuation is critical, the
Weyl fermions do not acquire a finite anomalous dimension in the
low-energy regime and the quasiparticle residue remains finite,
namely $Z_f \neq 0$. This indicates that the SC quantum fluctuation
does not qualitatively modify the low-energy properties of the
system, and that the fermions behave in nearly the same way as free
fermion gas in the non-SC phase. A simple mean-field treatment
should suffice to describe the transition. We thus obtain an example
of quantum phase transition that is characterized by trivial
critical phenomena. As illustrated by Fig.~1, a large NFL-like
quantum critical regime exists between the disordered and ordered
phases in many quantum critical systems. In contrast, there is not
such a NFL regime in 3D AWSM, which is caused by the special
anisotropy of fermion dispersion.

The rest of the paper is organized as follows. In
Sec.~\ref{Sec:MeanField}, we first make a mean-field analysis and
determine the SC QCP by solving the gap equation. In
Sec.~\ref{Sec:RGEQAnalysis}, we will go beyond the mean-field level
and study the influence of the quantum critical fluctuation of SC
order parameter by performing a RG analysis. The low-energy behavior
of all the model parameters are obtained from the solutions of the
self-consistent RG equations. We briefly summarize the results and
also discuss the possible experimental probe of our prediction in
Sec.~\ref{Sec:Summary}. The details of mean-field calculation and RG
calculation are presented in Appendix~\ref{Appendix:GapEquation} and
Appendix~\ref{Appendix:RGAnalysis}, respectively.

\section{Superconducting transition \label{Sec:MeanField}}

The system under consideration is described by the Hamiltonian $H =
H_{0} + H_{I}$, where
\begin{eqnarray}
H_{0} &=& \sum_{\mathbf{k}}\psi_{\mathbf{k}}^{\dag}
(c_{f}k_{x}\sigma_{1}+c_{f}k_{y}\sigma_{2}+Ak_{z}^{2}
\sigma_{3})\psi_{\mathbf{k}}, \\
H_{I} &=& -g\sum_{\mathbf{k},\mathbf{q}}
\psi_{\mathbf{k}}^{\dag}(-i\sigma_2) \psi_{-\mathbf{k}}^{\dag}
\psi_{\mathbf{q}}(i\sigma_{2})\psi_{\mathbf{-q}},
\end{eqnarray}
where the fermion field operator is defined as
$\psi_{\mathbf{k}}^{\dag} = (c_{\mathbf{k},\uparrow}^{\dag},
c_{\mathbf{k},\downarrow}^{\dag})$ to implement the spinor structure
and $\sigma_{1,2,3}$ are the standard Pauli matrices. The fermion
dispersion \cite{Yang13, Yang14A, Yang14B} has the form
$E_{f}=\pm\sqrt{c_{f}^{2}k_{\bot}^{2}+A^{2}k_{z}^{4}}$, where
$k_{\perp}^2=k_{x}^2+k_{y}^2$, and $c_{f}$ and $A$ are two
parameters introduced to characterize the energy dispersions within
$x$-$y$ plane and along $z$-axis, respectively. Here, we consider
one single specie of anisotropic Weyl fermions. The short-range
pairing interaction is described by $H_{I}$, where the coupling
constant $g
> 0$.

We first make a mean-field analysis to determine the SC QCP. The SC
order parameter is defined as
\begin{eqnarray}
\Delta_{s} = g\sum_{\mathbf{k}}\langle
\psi_{\mathbf{k}}(i\sigma_{2})\psi_{\mathbf{-k}}\rangle.
\end{eqnarray}
At the mean-field level, we have
\begin{eqnarray}
H_{I} = \sum_{\mathbf{k}} \left[- \Delta_{s}^{*}
\psi_{\mathbf{k}}(i\sigma_{2})\psi_{\mathbf{-k}} + \Delta_{s}
\psi_{\mathbf{-k}}^{\dag}(i\sigma_2)\psi_{\mathbf{k}}^{\dag}\right]
+ \frac{1}{2g}|\Delta_{s}|^2, \nonumber
\end{eqnarray}
where the SC gap is supposed to be $s$-wave. According to the
calculations presented in Appendix A, the zero-temperature gap
equation is
\begin{eqnarray}
2\int\frac{dw}{2\pi}\int \frac{d^{3}\mathbf{k}}{(2\pi)^{3}}
\frac{1}{\omega^{2}+c_{f}^{2}k_{\bot}^{2}+A^{2}k_{z}^{4} +
|\Delta_{s}|^{2}} = \frac{1}{2g}.
\end{eqnarray}
It is easy to verify that a nonzero SC gap is opened only when the
coupling $g$ exceeds the critical value
\begin{eqnarray}
g_{c} = \frac{3(2\pi)^2 c_f^2\sqrt{A}}{4E_{D}^{3/2}},
\end{eqnarray}
where $E_{D}$ is a cutoff.

\section{Renormalization group study of quantum critical behavior \label{Sec:RGEQAnalysis}}

At the SM-SC QCP, the SC order parameter vanishes, namely $\langle
\psi_{\mathbf{k}}(i\sigma_{2})\psi_{\mathbf{-k}}\rangle = 0$. But
its quantum fluctuation cannot be simply neglected. We will carry
out a RG analysis to examine whether or not its quantum fluctuation
leads to significant effects on the low-energy behavior of
anisotropic Weyl fermions.

The quantum critical system can be modeled by the following
effective action
\begin{eqnarray}
S = S_{\psi} + S_{\phi} + S_{\phi^4} + S_{\psi\phi},
\end{eqnarray}
where the free action for Weyl fermions is given by
\begin{eqnarray}
S_{\psi} = \int \frac{d\omega}{2\pi}
\frac{d^{3}\mathbf{k}}{(2\pi)^{3}} \psi^{\dagger}
\left[-i\omega\sigma_{0}+\mathcal{H}_{\psi}(\mathbf{k})\right]\psi,
\end{eqnarray}
where
$\mathcal{H}_{\psi}(\mathbf{k})=c_{f}k_{x}\sigma_{1}+c_{f}k_{y}\sigma_{2}
+ Ak^{2}_{z}\sigma_{3}$, the one for SC order parameter is
\begin{eqnarray}
S_{\phi} = \frac{1}{2}\int\frac{d\omega}{2\pi}
\frac{d^{3}\mathbf{q}}{(2\pi)^{3}}\phi^{*}
\left[\Omega^{2}+E_{\phi}^{2}(\mathbf{q}) + r\right]\phi,
\end{eqnarray}
where $E_{\phi}(\mathbf{q})=\sqrt{c_{b\perp}^{2}\left(q_{x}^{2} +
q_{y}^{2}\right) + c_{bz}^{2}q_{z}^{2}}$. Here, $c_{b\perp}$ is the
boson velocity within the $x$-$y$ plane and $c_{bz}$ the one along
$z$-direction. The boson mass $r$ serves as a tuning parameter: $r >
0$ corresponds to SM phase ($g < g_c$) and $r < 0$ to SC phase ($g >
g_c$). In the following, we focus on the SM-SC QCP, corrsponding to
$r=0$. The free fermion and boson propagators are
\begin{eqnarray}
G_{\psi}(\omega,\mathbf{k}) &=& \frac{1}{-i\omega\sigma_{0} +
c_{f}k_{x}\sigma_{1} + c_{f}k_{y}\sigma_{2} +
Ak^{2}_{z}\sigma_{3}},\\
G_{\phi}(\Omega,\mathbf{q}) &=& \frac{1}{\Omega^{2} +
c_{b\perp}^{2}q_{x}^{2} + c_{b\perp}^{2}q_{y}^{2} +
c_{bz}^{2}q_{z}^{2}}.
\end{eqnarray}

The self-coupling of the boson field takes the form
\begin{eqnarray}
S_{\phi^{4}} = \frac{\lambda}{4}\int\prod_{i=1}^{4}
\frac{d\Omega_{i}}{2\pi}\frac{d^{3}\mathbf{q}_{i}}{(2\pi)^{3}}
D(\Omega)D(\mathbf{q})|\phi|^{4}, \label{Eq:ActionFourBoson}
\end{eqnarray}
where for simplicity we define
\begin{eqnarray}
D(\Omega) &\equiv& \delta(\Omega_{1}+\Omega_{3} -
\Omega_{2}-\Omega_{4}),\nonumber \\
D(\mathbf{q}) &\equiv& \delta^{3}(\mathbf{q_{1}} +
\mathbf{q_{3}}-\mathbf{q_{2}}-\mathbf{q_{4}}).
\end{eqnarray}
The Yukawa-coupling between the gapless fermions and the critical
boson is described by
\begin{eqnarray}
S_{\psi\phi} &=& h\int\prod_{i=1}^{2} \frac{d\omega_{i}}{2\pi}
\frac{d^{3}\mathbf{k}_{i}}{(2\pi)^{3}}\frac{d\Omega}{2\pi}
\frac{d^{3}\mathbf{q}}{(2\pi)^{3}}\delta(\omega_{1}+\omega_{2}-\Omega)
\nonumber \\
&&\times\delta^{3}(\mathbf{k_{1}}+\mathbf{k_{2}}-\mathbf{q})
(\phi^{*}\psi^{T}i\sigma_{2}\psi+H.c.),
\end{eqnarray}
where $h$ is the coupling constant.

\begin{figure}[htbp]
\center
\includegraphics[width=1.65in]{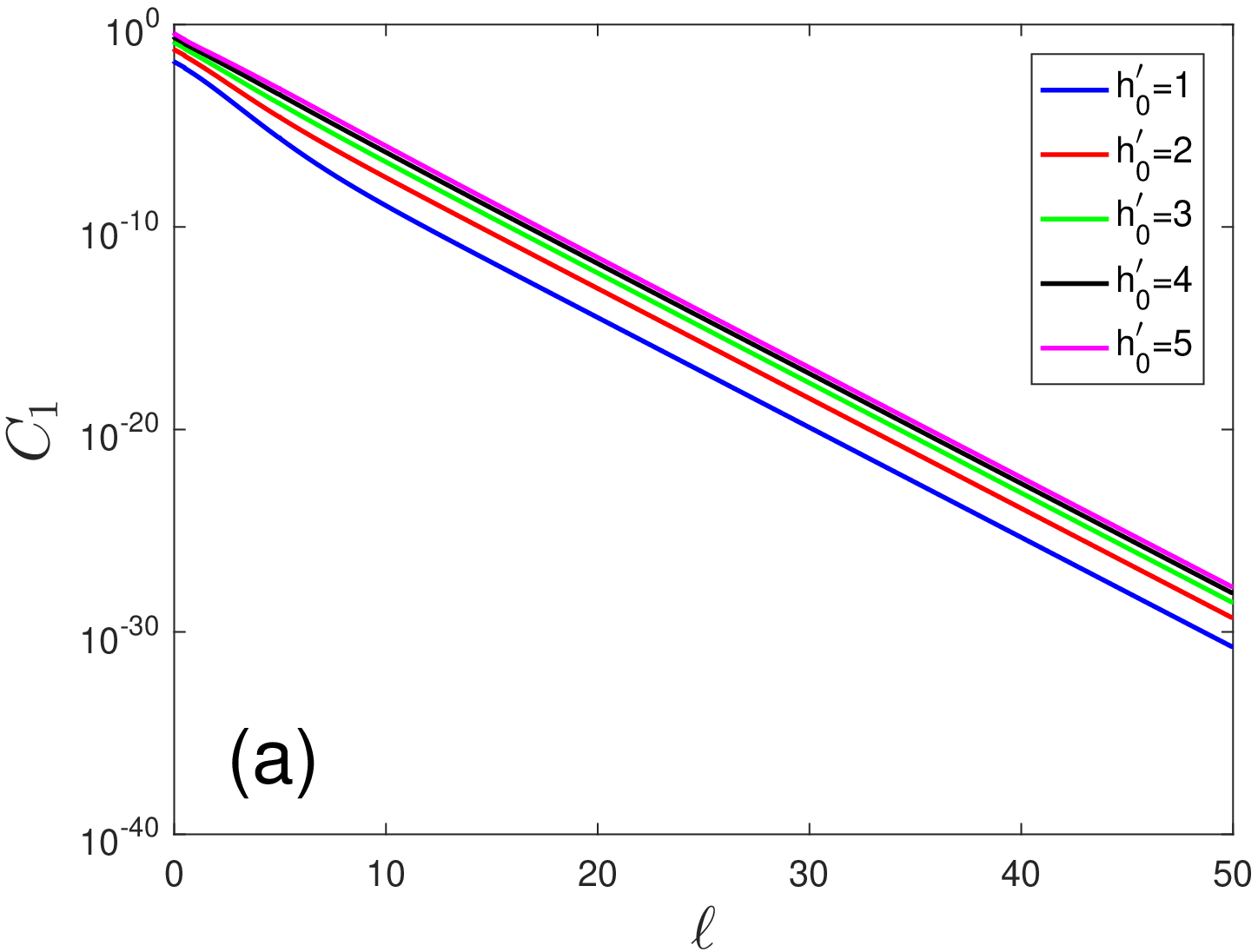}
\includegraphics[width=1.65in]{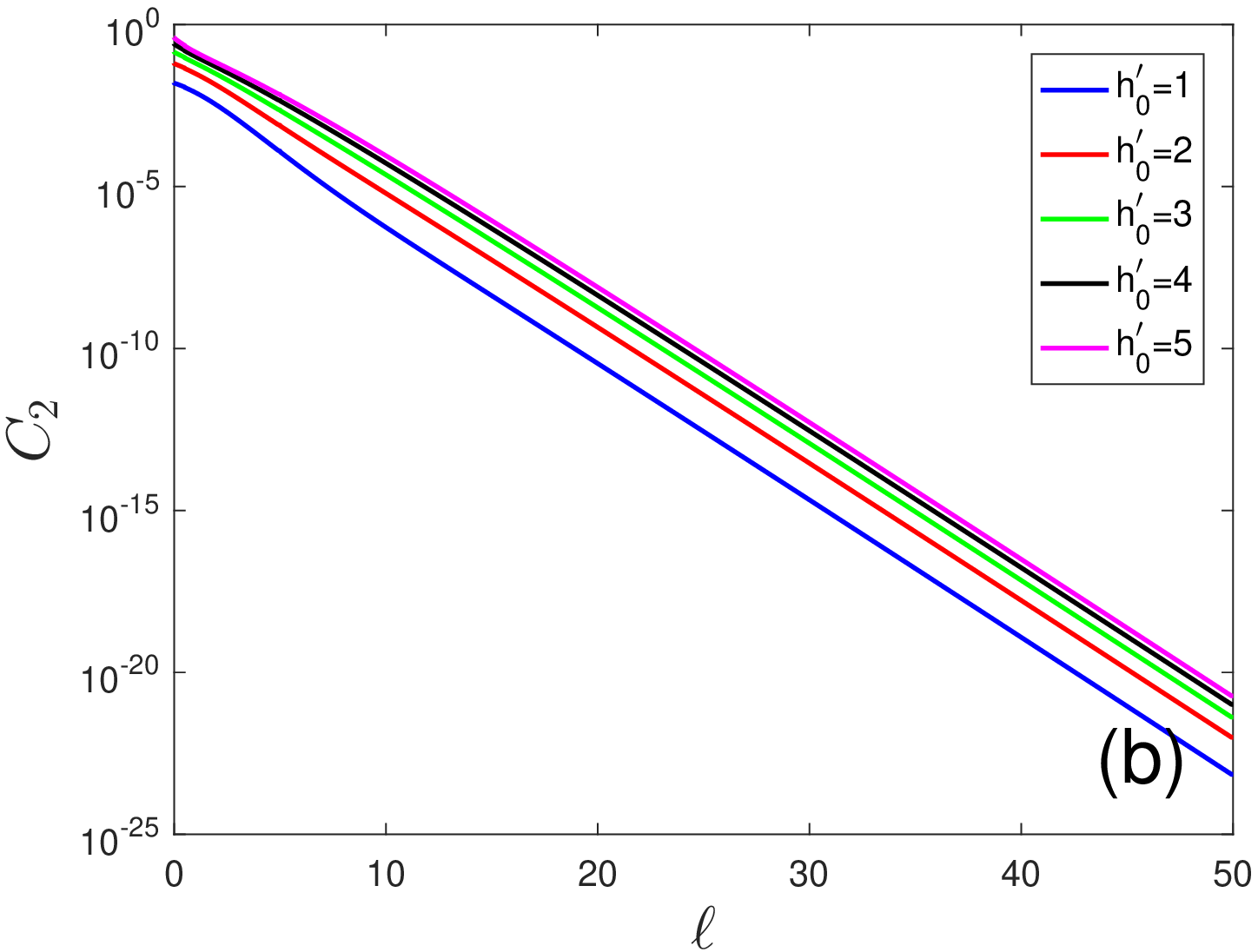}
\includegraphics[width=1.65in]{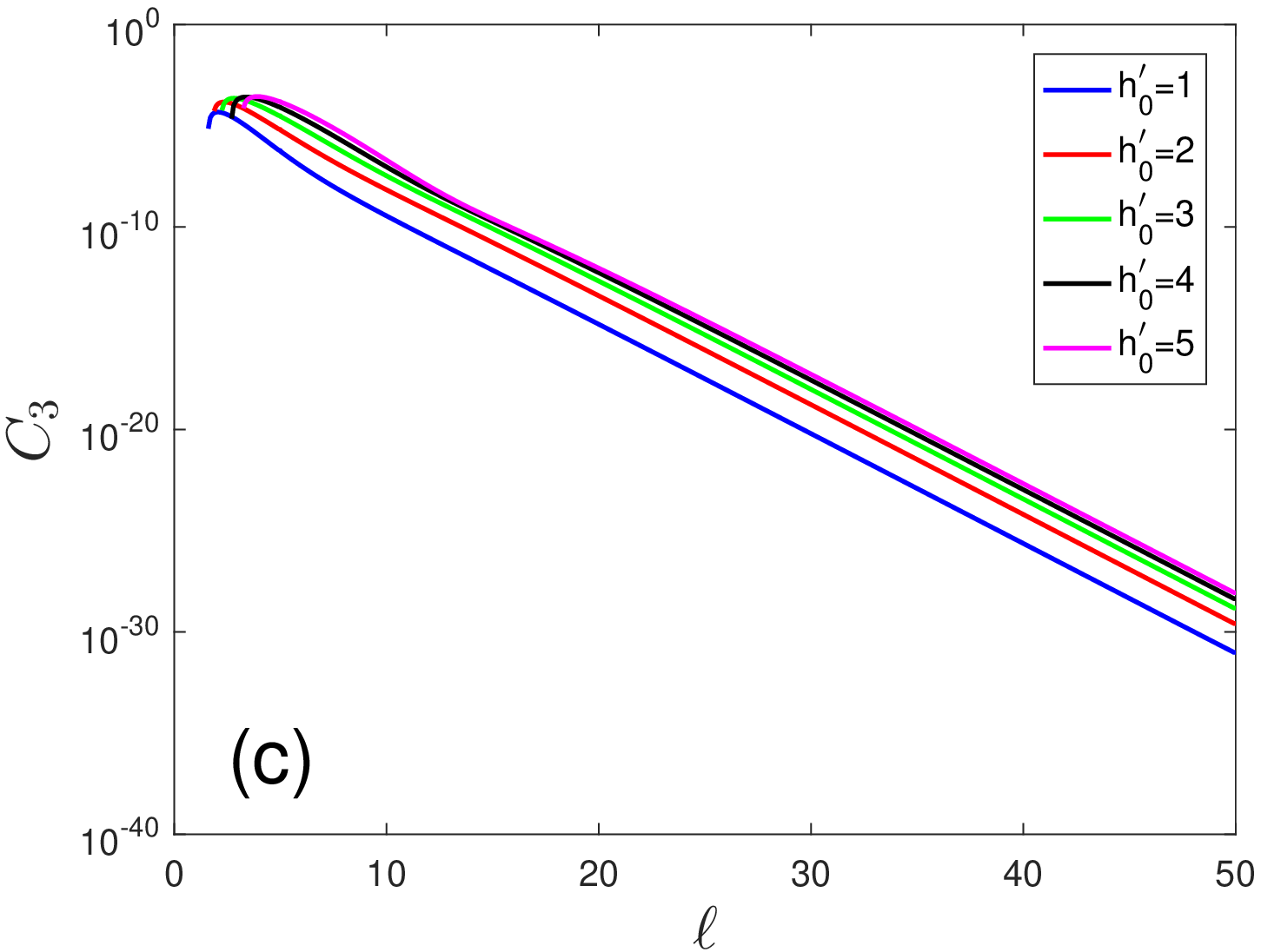}
\includegraphics[width=1.65in]{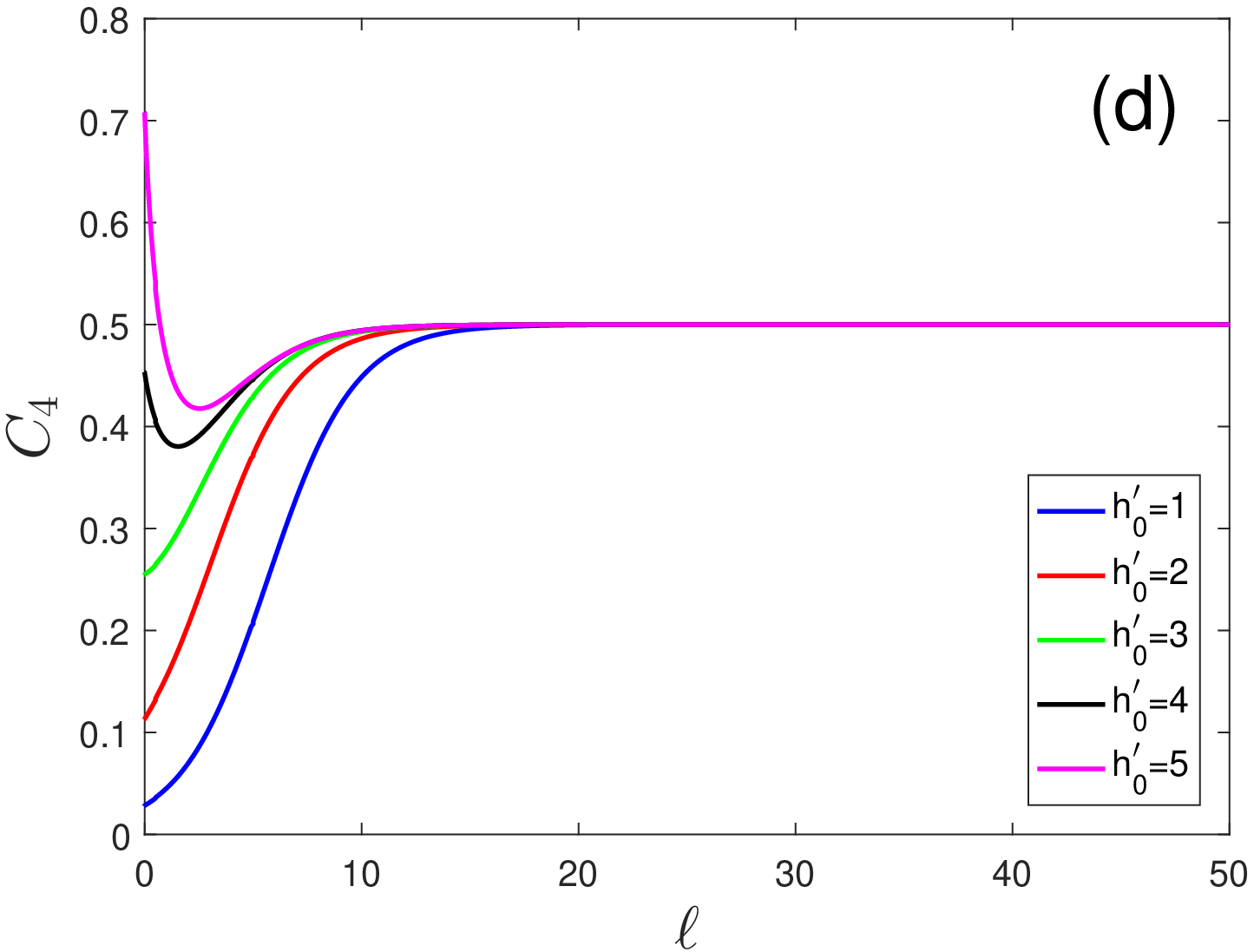}
\includegraphics[width=1.65in]{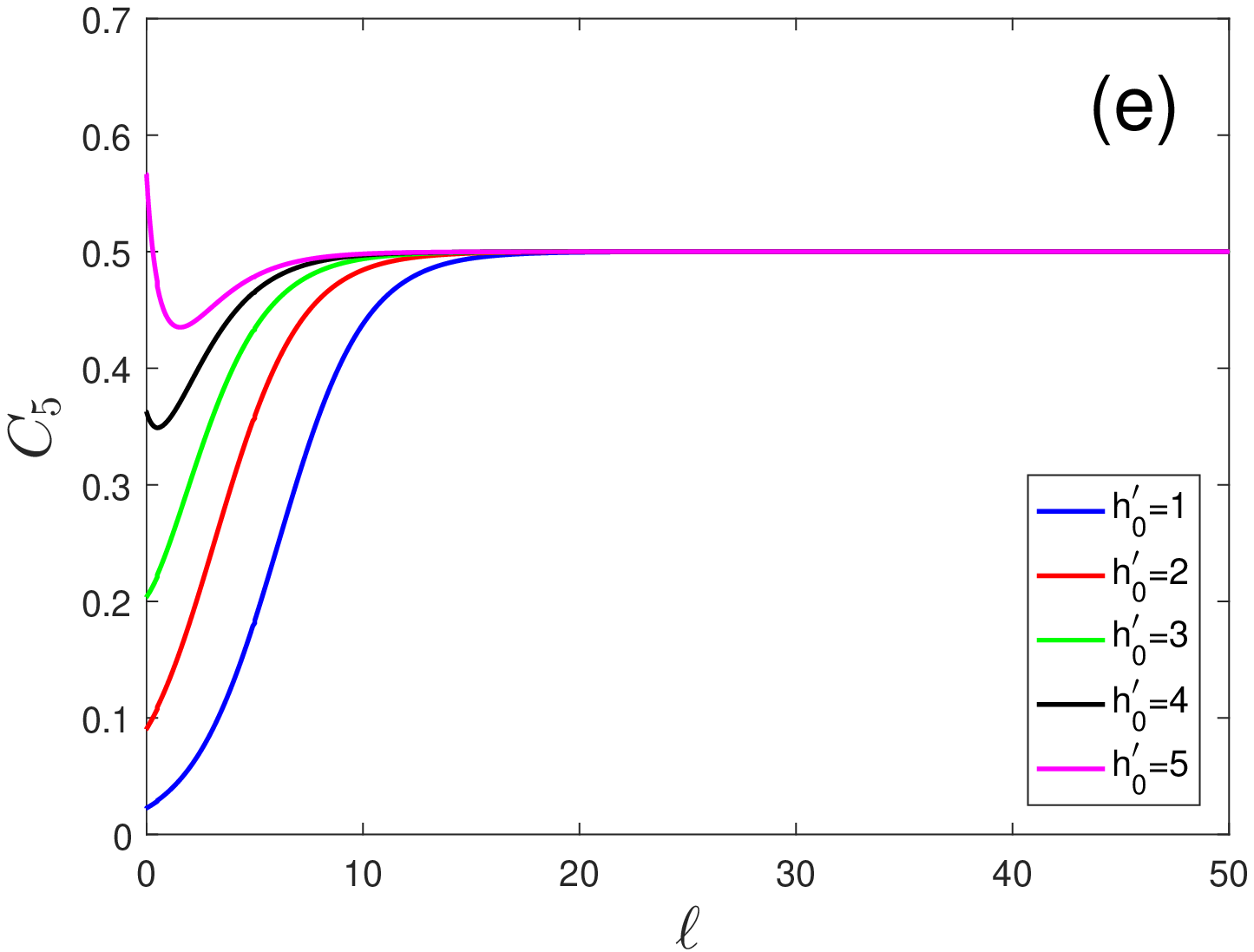}
\includegraphics[width=1.65in]{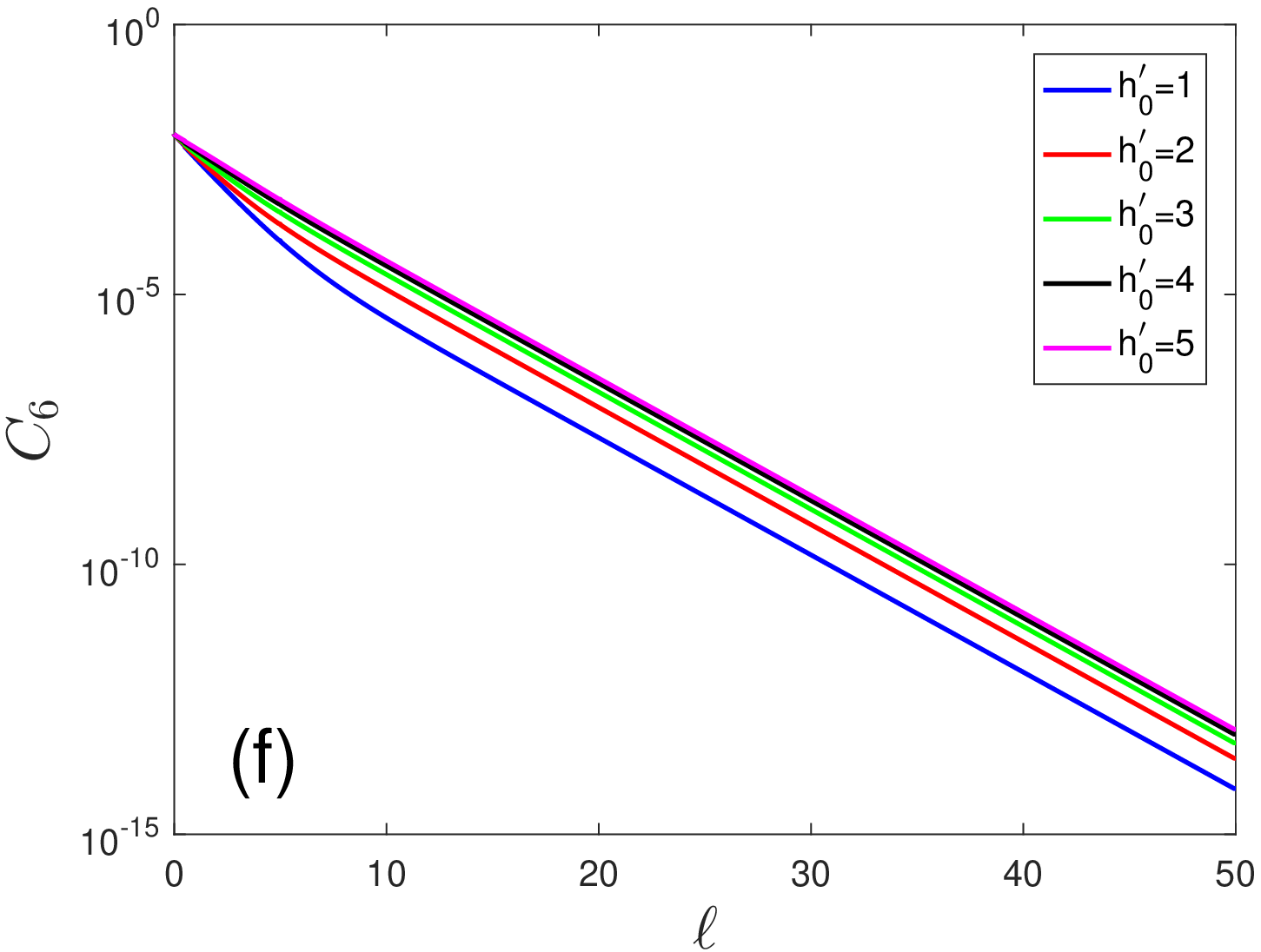}
\includegraphics[width=1.65in]{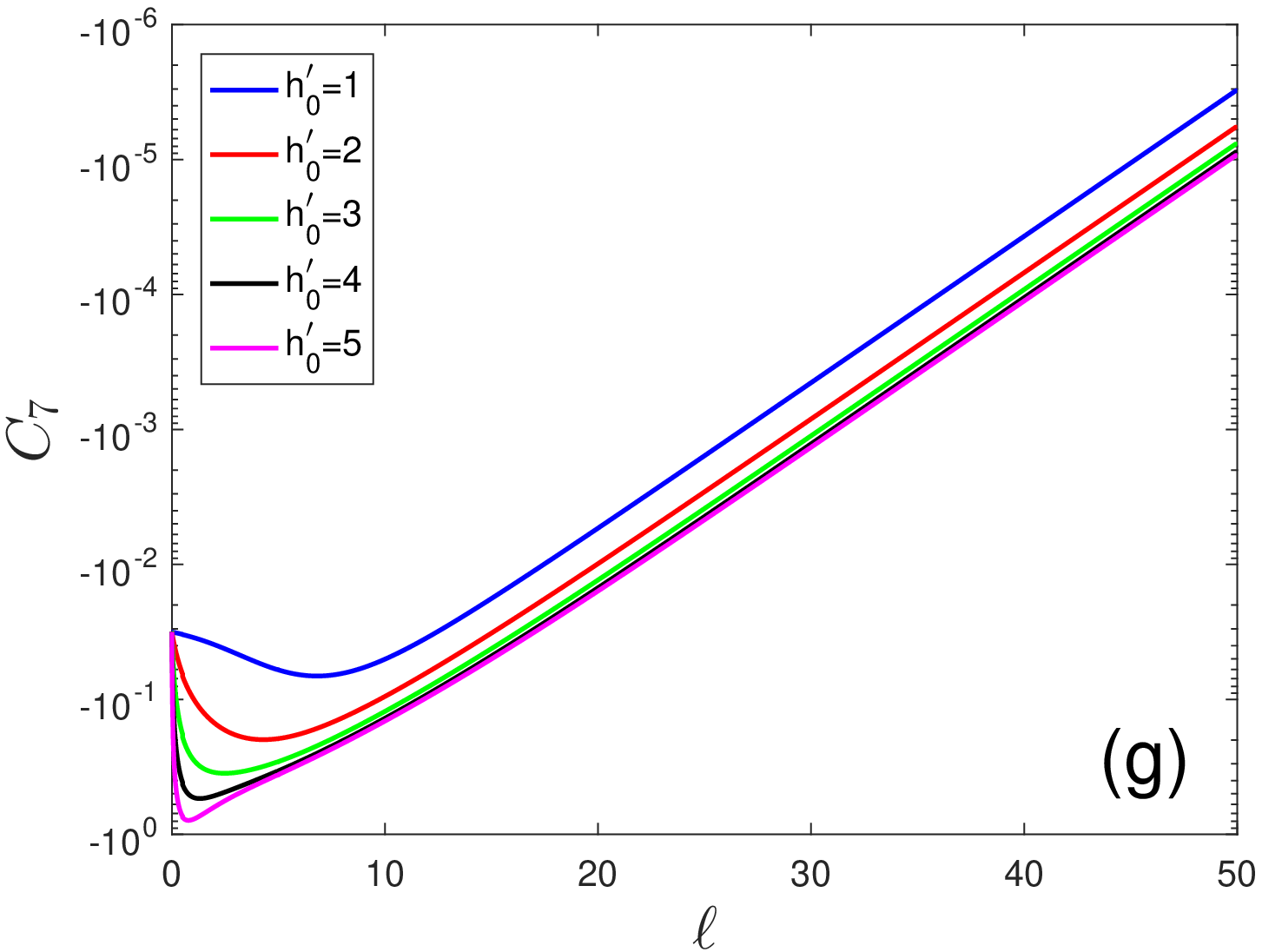}
\includegraphics[width=1.65in]{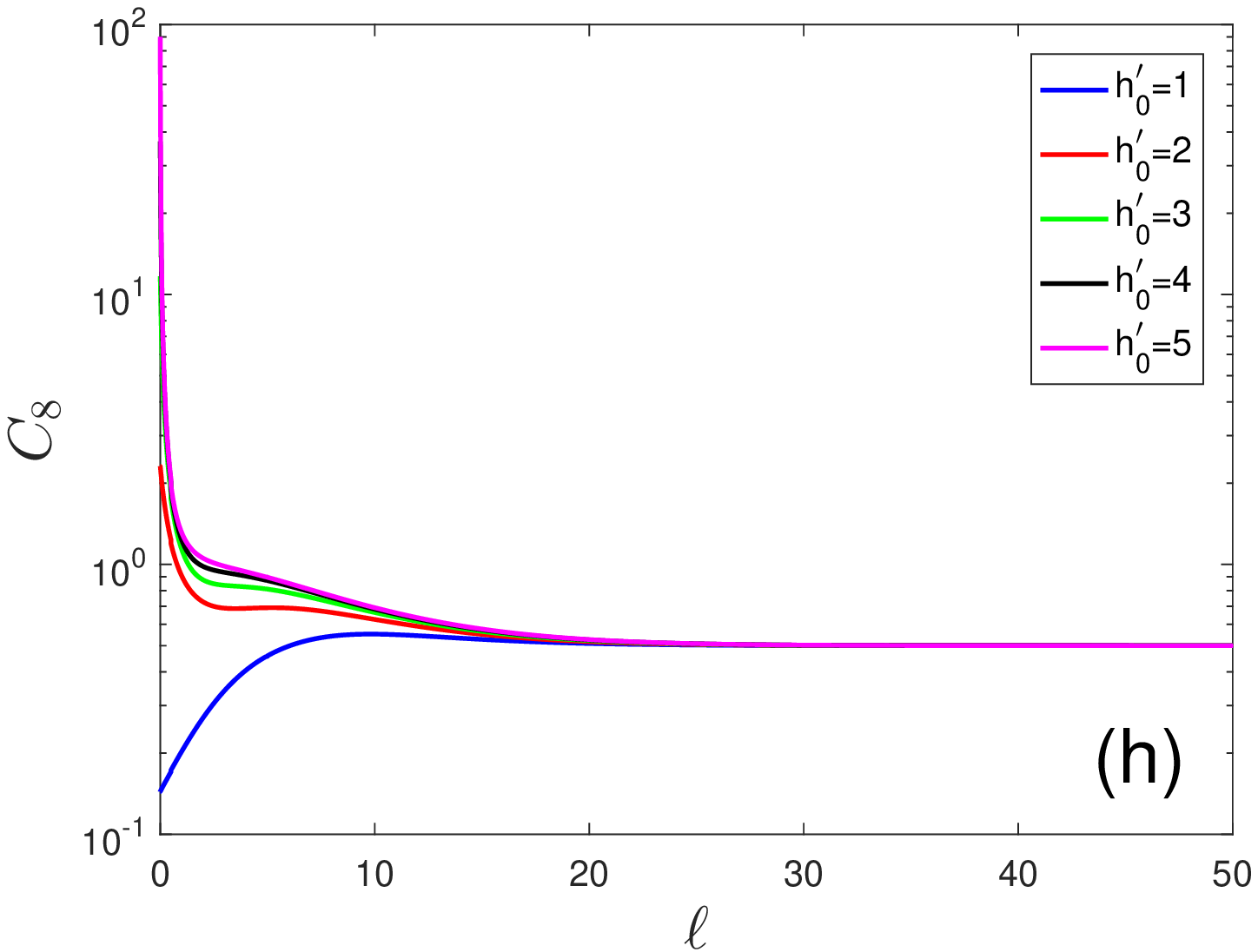}
\caption{Scale dependence of constants $C_{i}$ ($i=1,2,...,8$) at
different initial values of $h_{0}'$. Here, we suppose that
$\lambda'_{0}=0.5$, $\varsigma_{0}=0.2$, $\eta_{A0}=1$, and
$\eta_{B0}=1$, which will be used in all the following calculations.
\label{Fig:VRGCi}}
\end{figure}

The whole action contains six model parameters, namely $c_f$, $A$,
$c_{b\bot}$, $c_{bz}$, $\lambda$, and $h$. These parameters all
receive quantum corrections from the Yukawa coupling, and then
become scale dependent. The low-energy critical behavior of the SC
QCP can be analyzed based on the scale dependence of all these
parameters. After carrying out lengthy calculations, with full
details presented in Appendices B and C, we derive the following
coupled RG equations:
\begin{eqnarray}
\frac{d{c_{f}}}{d\ell}&=&\left(-C_{1}+C_{2}\right)c_{f}, \label{Eq:VRGcf} \\
\frac{dA}{d\ell}&=&\left(-C_{1}+C_{3}\right)A, \label{Eq:VRGA} \\
\frac{dc_{b\perp}}{d\ell}&=&\frac{1}{2}\left(-C_{4}+C_{5}\right)c_{b\perp},
\label{Eq:VRGcbbot} \\
\frac{dc_{bz}}{d\ell}&=&\frac{1}{2}\left(1-C_{4}+C_{6}\right)c_{bz},
\label{Eq:VRGcbz} \\
\frac{d\lambda}{d\ell}&=&\left(\frac{1}{2}-2C_{4}+C_{7}+C_{8}\right)\lambda,
\label{Eq:VRGlambda} \\
\frac{dh}{d\ell}&=&\left(\frac{1}{4}-C_{1}-\frac{C_{4}}{2}\right)h.
\label{Eq:VRGh}
\end{eqnarray}
Here, $\ell$ is a freely varying scale, and the lowest energy limit
corresponds to $\ell \rightarrow \infty$. The analytical expressions
of $C_i$, with $i=1,2,..., 8$, are given in Appendix B. The
$\ell$-dependence of $C_i$ can be obtained by numerically solving
these equations. The solutions are shown in Fig.~\ref{Fig:VRGCi}.

\begin{figure}[htbp]
\center
\includegraphics[width=2.6in]{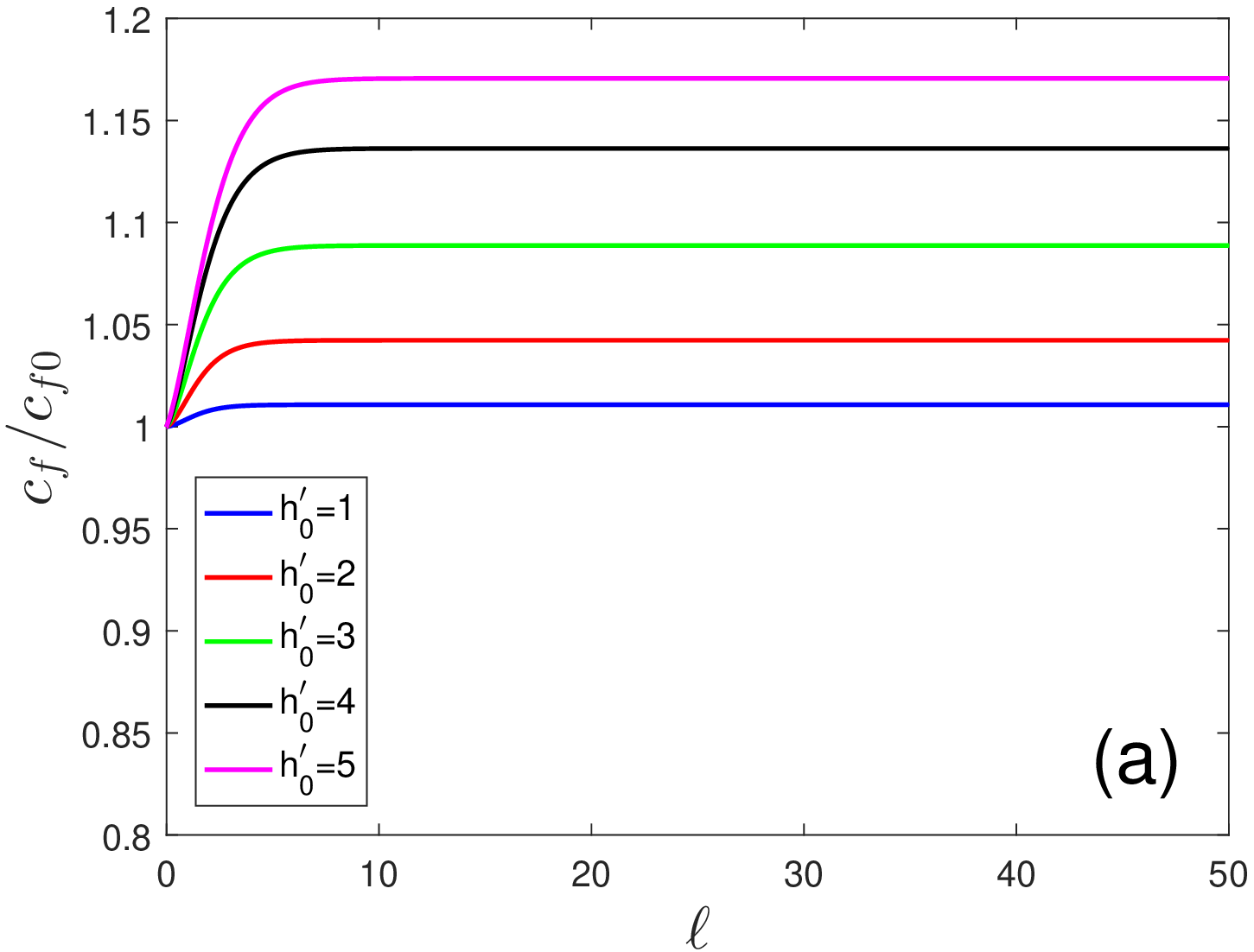}
\includegraphics[width=2.6in]{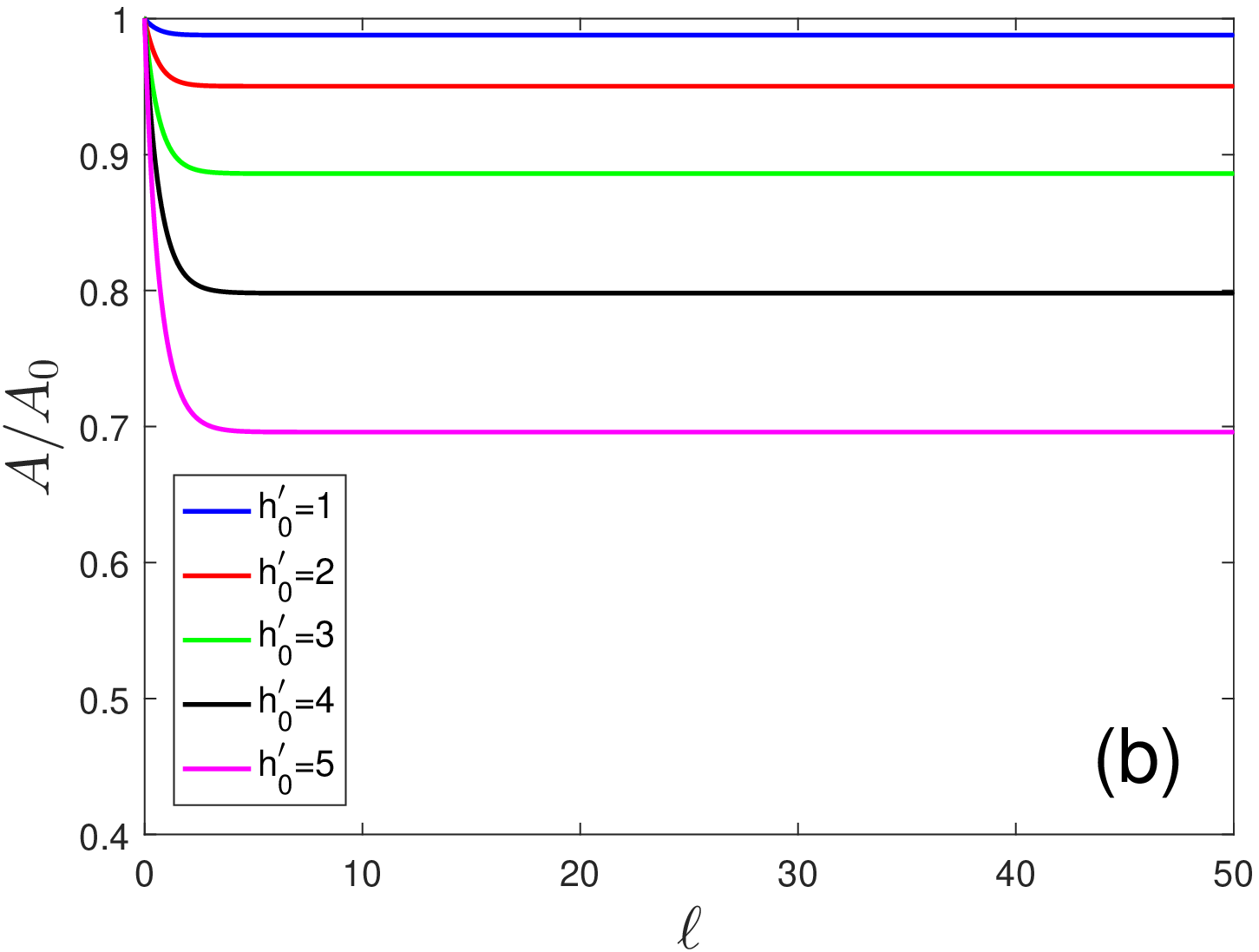}
\caption{Scale dependence of $v$ and $A$ at different initial values
of $h_{0}'$. \label{Fig:VRGCfA}}
\end{figure}

To examine the impact of interactions, it is convenient to define
two new parameters
\begin{eqnarray}
\lambda' = \lambda/c_{f}^{3}, \quad h' = h/c_{f}^{3/2}.
\end{eqnarray}
Now we can re-write the equations for $\lambda'$ and $h'$ as
\begin{eqnarray}
\frac{d\lambda'}{d\ell} &=& \left(\frac{1}{2}+3C_{1}-3C_{2}-2C_{4} +
C_{7}+C_{8}\right)\lambda', \label{Eq:VRGlambdaPrime}
\\
\frac{dh'}{d\ell} &=& \left(\frac{1}{4}+ \frac{1}{2}C_{1} -
\frac{3}{2}C_{2}-\frac{C_{4}}{2}\right)h'. \label{Eq:VRGhPrime}
\end{eqnarray}

The dependence of $c_{f}$ and $A$ on the running energy scale $\ell$
is shown in Fig.~\ref{Fig:VRGCfA}. We can find that $c_f$ and $A$
are only quantitatively modified, and approach to new constant
values. The indication is that, the observable quantities, such as
fermion DOS and specific heat, exhibit nearly the same behavior as
the free fermion system.

\begin{figure}[htbp]
\center
\includegraphics[width=2.6in]{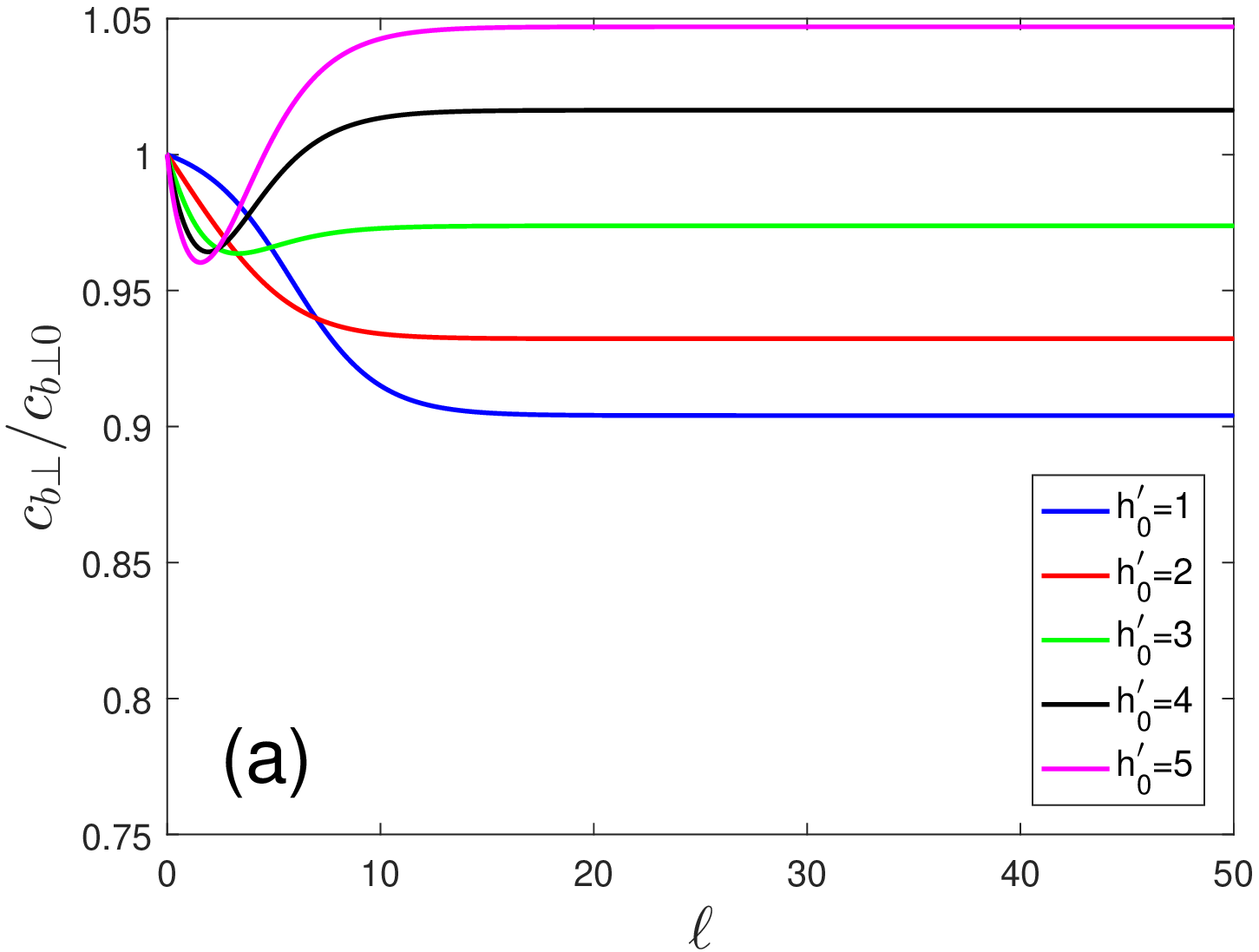}
\includegraphics[width=2.6in]{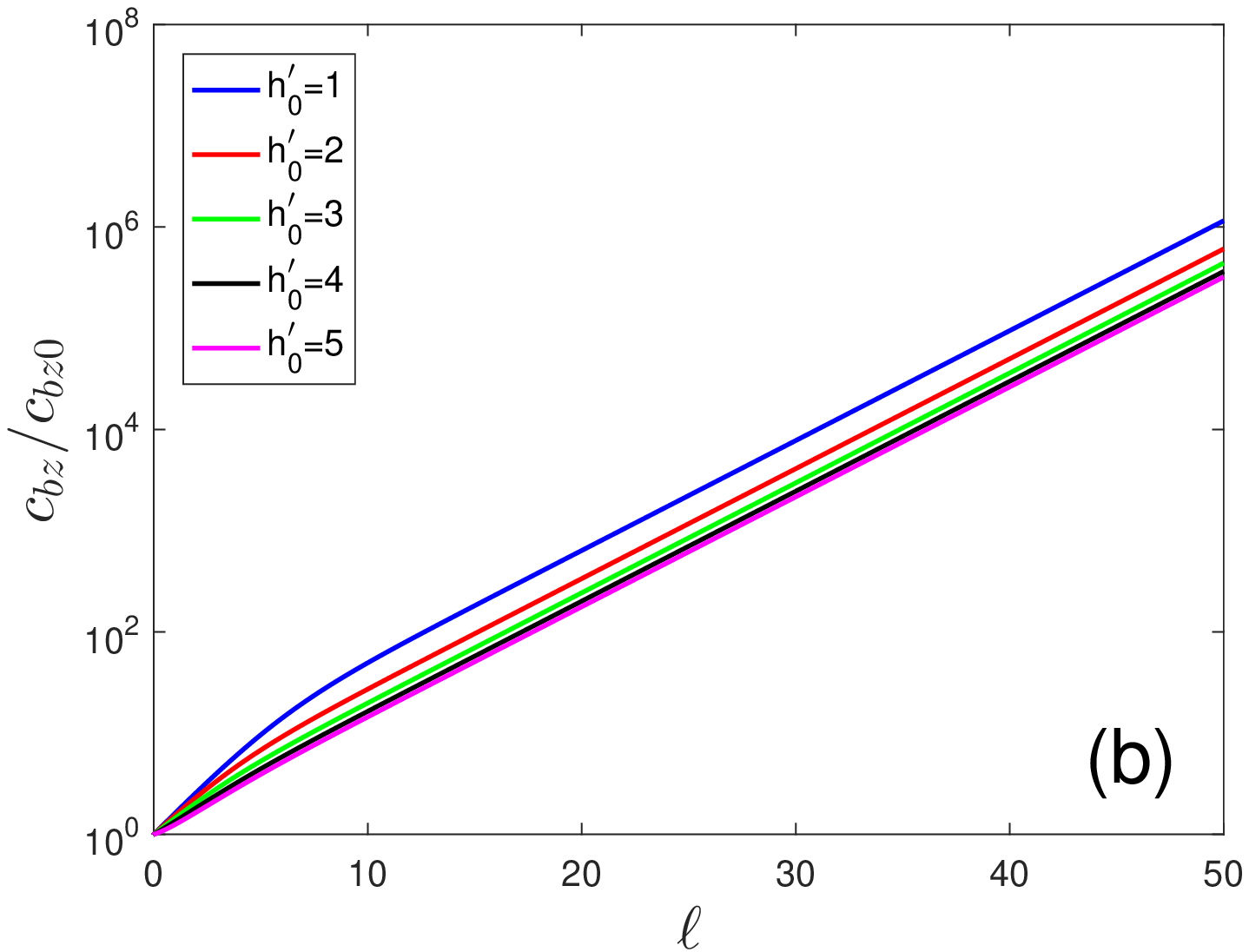}
\caption{Scale dependence of $c_{b\bot}$ and $c_{bz}$ at different
initial values of $h_{0}'$. \label{Fig:VRGCbBotZ}}
\end{figure}

As can be seen from Fig.~\ref{Fig:VRGCbBotZ}(a), the parameter
$c_{b\bot}$ flows to a different constant value in the lowest energy
limit. According to Eq.~(\ref{Eq:VRGcbz}), $c_{bz}$ flows to
infinity even when one-loop corrections are not included. This
results from the property that the momentum component along $z$
direction scales differently from the components within $x$-$y$
plane. After including one-loop corrections, $c_{bz}$ still flows to
infinity, but at a lower speed, as shown in
Fig.~\ref{Fig:VRGCbBotZ}(b).

\begin{figure}[htbp]
\center
\includegraphics[width=2.6in]{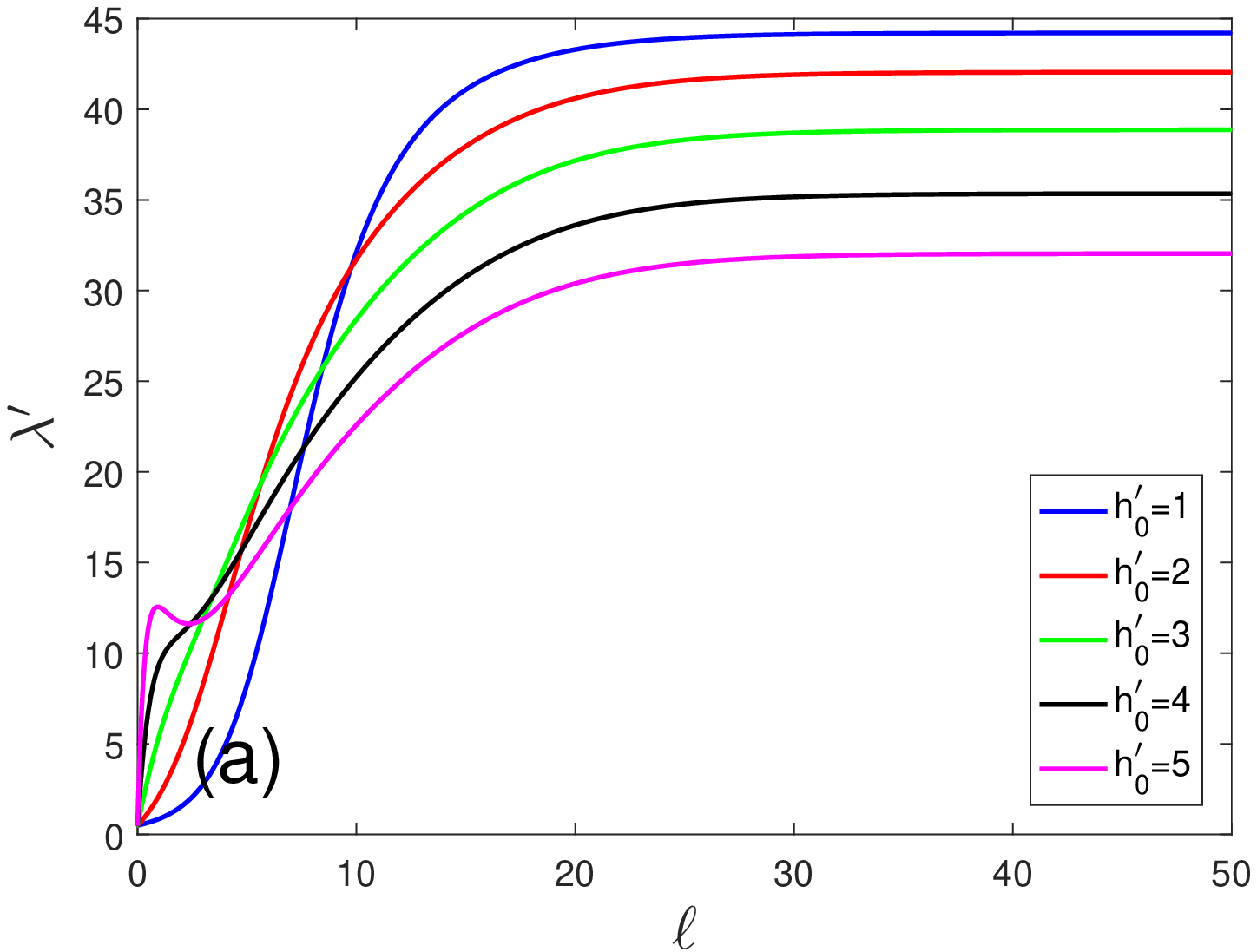}
\includegraphics[width=2.6in]{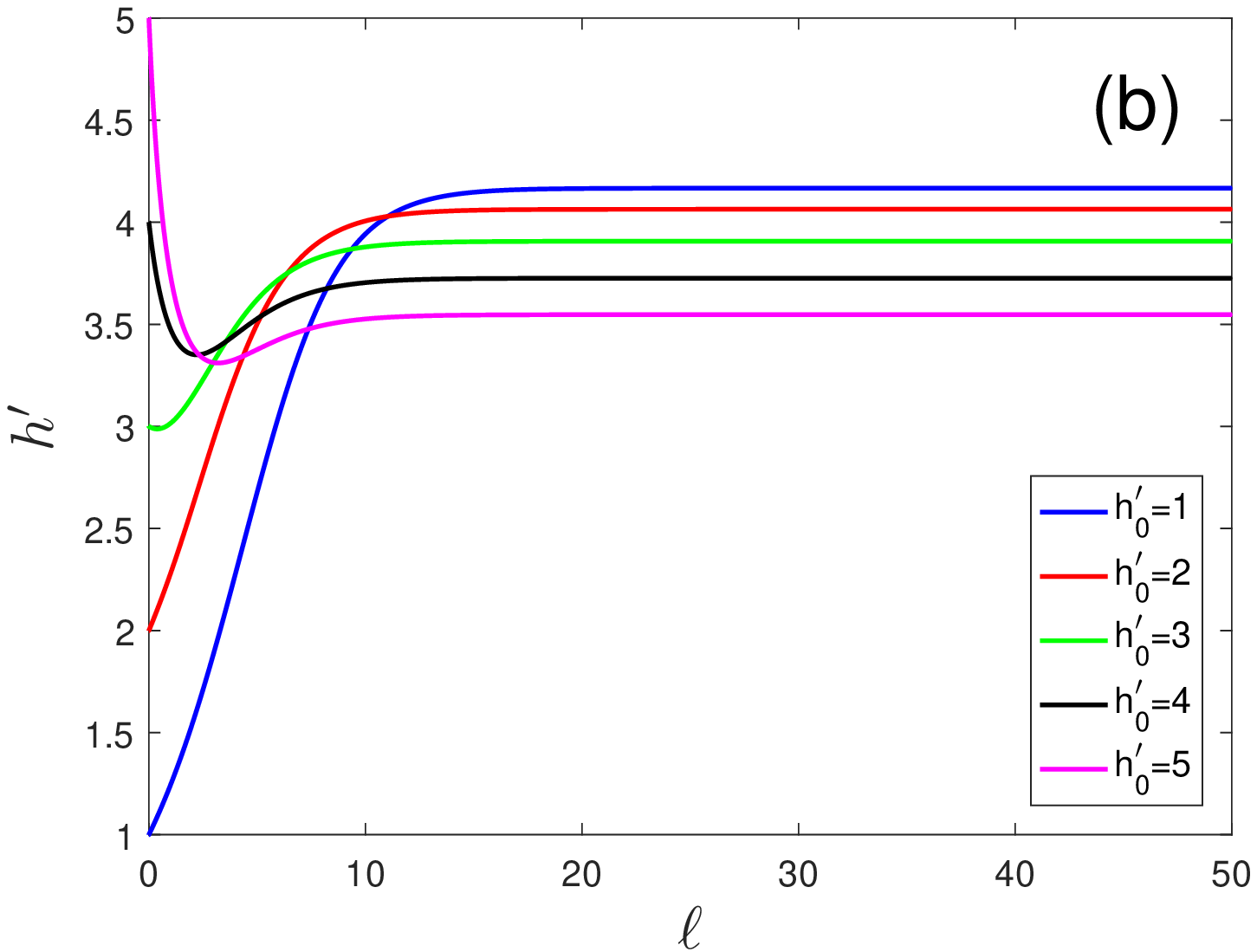}
\caption{Flows of $\lambda'$ and $h'$ at different initial values of
$h_{0}'$. \label{Fig:VRGClambdah}}
\end{figure}

\begin{figure}[htbp]
\center
\includegraphics[width=2.8in]{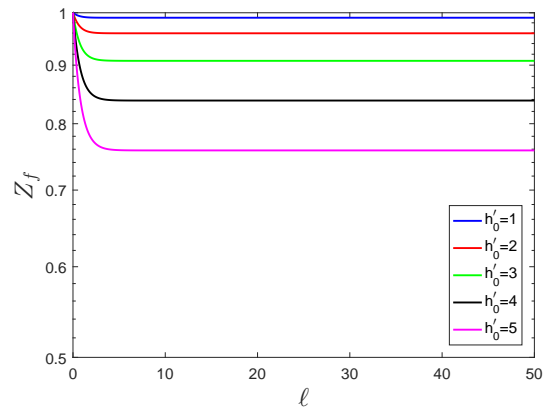}
\caption{The quasiparticle residue $Z_f$ flows to a constant at
large $\ell$, independent of the value of $h_0'$. \label{Fig:VRGZf}}
\end{figure}

We present the $\ell$-dependence of coupling constants $\lambda'$
and $h'$ in Fig.~\ref{Fig:VRGClambdah}. Both of $\lambda'$ and $h'$
flow to certain finite constants in the lowest energy limit, namely
$\lambda'\rightarrow \lambda'^{\ast}$ and $h'\rightarrow h'^{\ast}$.
Thus, $\lambda'$ and $h'$ are both marginal. The values of
$\lambda'^{\ast}$ and $h'^{\ast}$ depend on the bare values of
$\lambda'$ and $h'$. The flowing behavior of $C_{i}$ with
$i=1,2,..,8$ in Figs.~\ref{Fig:VRGCi}(a)-(h), respectively.
According to Figs.~\ref{Fig:VRGCi}(a), (b), and (c), we observe that
$C_{1}$, $C_{2}$, and $C_{3}$ flow to zero very quickly. As a
result, the parameters $c_f$ and $A$ do not receive singular
renormalization, but flow to finite constants. The anomalous
dimension of fermion field is given by $\eta_{f} = C_{1}$. Since
$C_{1}$ vanishes rapidly at low energies, we infer that the fermion
field does not acquire any anomalous dimension. The flow equation of
quasiparticles residue $Z_f$ is
\begin{eqnarray}
\frac{dZ_{f}}{d\ell} = -C_{1}Z_{f}.
\end{eqnarray}
As shown in Fig.~(\ref{Fig:VRGZf}), $Z_{f}$ always flows to a finite
constant in the lowest energy limit. These results indicate that the
anisotropic Weyl fermions are well-defined quasiparticles and have a
long lifetime at the SM-SC QCP.

We now analyze the impact of Yukawa coupling on the bosonic mode.
Figs.~\ref{Fig:VRGCi}(d)-(f) show that
\begin{eqnarray}
C_{4} \rightarrow 0.5, \quad C_{5} \rightarrow 0.5, \quad C_{6}
\rightarrow 0
\end{eqnarray}
in the lowest energy limit. The RG equation for parameter $c_{bz}$
becomes
\begin{eqnarray}
\frac{dc_{bz}}{d\ell} = \frac{1}{2}(1-C_{4}+C_{6})c_{bz} \approx
0.25 c_{bz}.
\end{eqnarray}
The one-loop order correction does not lead to qualitative change of
the flow of $c_{b\bot}$. Therefore, the bosonic SC fluctuation is
anisotropically screened. Since $C_{4}$ flows to a finite value at
low energies, the boson field $\phi$ acquires a finite anomalous
dimension. Now the renormalized boson propagator becomes
\begin{eqnarray}
G_{\phi}(\Omega,\mathbf{q})\sim \frac{1}{\left(\Omega^{2} +
c_{b\bot}^{2}q_{\bot}^{2}\right)^{3/4}+c_{bz}^{2}q_{z}^{2}},
\end{eqnarray}
where $q_{\perp}^2 = q_{x}^2+q_{y}^2$.

According to Figs.~\ref{Fig:VRGCi}(g) and (h), we see that $C_{7}
\rightarrow 0$ and $C_{8}\rightarrow 0.5$ in the lowest energy
limit. Combining Eq.~(\ref{Eq:VRGlambdaPrime}),
Eq.~(\ref{Eq:VRGhPrime}), and the low-energy behavior of $C_{1}$,
$C_{4}$, $C_{7}$, and $C_{8}$, we conclude that the beta functions
of $\lambda'$ and $h'$ vanish, which explains why both $\lambda'$
and $h'$ approach finite constants.

To understand the peculiarity of our result, we now compare it to
previous studies of various quantum critical systems.
Superconductivity was proposed to occur in several SM materials,
including 2D Dirac SM \cite{Lee07, Ponte14, Grover14,
Witczak-Krempa16, Zerf16}, Luttinger SM \cite{Boettcher16}, and 3D
Weyl SM \cite{Jian15}. In 2D Dirac SM and Luttinger SM, the system
flows to a stable infrared fixed point at the SC QCP. At such a
fixed point, the fermion field acquires a finite anomalous
dimension, which leads to power-law correction to the fermion DOS.
Moreover, the fermion damping rate behaves as $\Gamma(\omega)\propto
|\omega|^{1-\eta_{f}}$ at low energies, and the residue $Z_f\sim
|\omega|^{\eta_{f}}\rightarrow 0$ in the limit $\omega \rightarrow
0$. In the case of 3D Weyl SM, the anomalous dimension of fermion
field approaches to zero very slowly at the SC QCP \cite{Jian15}.
Then, the fermion DOS receives logarithmic-like correction.
Accordingly, the residue $Z_f$ also flows to zero very slowly, and
the fermion damping rate exhibits marginal Fermi liquid behavior. We
thus see that FL theory breaks down at the SM-SC QCP in all these
systems. The singular fermion DOS revealed in previous theoretical
works can be verified by scanning tunneling microscope (STM)
measurements. In addition, angle resolved photoemission spectroscopy
(ARPES) experiments \cite{Damascelli03} may be applied to probe the
strong NFL behavior.

Similar NFL-like quantum critical phenomena also emerge in metals
that are tuned to the vicinity of a continuous quantum phase
transition. It is well-established that NFL behavior is realized
near the FM, AFM, and nematic QCPs. For instance, the zero-$T$
Landau damping rate is $\Gamma(\omega)\propto |\omega|^{1/2}$ at an
AFM QCP \cite{Abanov03, Metlitski10AFM} and $\Gamma(\omega)\propto
|\omega|^{2/3}$ at an FM or Ising-type nematic QCPs \cite{Rech06,
Metzner03}. The corresponding residue $Z_f = 0$ at all of these
QCPs.

Different from the above quantum critical systems, the fermion
anomalous dimension flows to zero very quickly and the residue $Z_f
\neq 0$ at the SC QCP in 3D AWSM. It thus turns out that such QCP
exhibits a trivial quantum criticality. Nevertheless, it is the
triviality that makes this system distinctive. As shown in
Fig.~1(a), a finite NFL regime exists on the phase diagram of
conventional quantum critical systems. There is no such NFL regime
in the system considered in this work. We observe from Fig.~1(b)
that, although there is a clear QCP between the gapless SM phase and
gapped SC phase, the anisotropic Weyl fermions do not display NFL
behavior around this QCP. The vanishing of fermion anomalous
dimension is closely related to the unusual anisotropic screening of
the SC quantum fluctuation, which in turn is induced by the special
dispersion of anisotropic Weyl fermions. Therefore, it is the strong
anisotropy of fermion dispersion that distinguishes the 3D AWSM from
all the other quantum critical systems. Indeed, if the fermion
dispersion took a different form, the SC quantum fluctuation might
lead to NFL-like quantum critical phenomena.

The trivial quantum criticality can be probed by measuring some
observable quantities. In the non-interacting limit, the fermion DOS
depends on energy as
\begin{eqnarray}
\rho(\omega)\propto \frac{|\omega|^{3/2}}{c_{f}^{2}\sqrt{A}},
\end{eqnarray}
and the specific heat depends on $T$ as
\begin{eqnarray}
C_{V}(T)\propto \frac{T^{5/2}}{c_{f}^{2}\sqrt{A}}.
\end{eqnarray}
Since $c_f$ and $A$ are not singularly renormalized, both
$\rho(\omega)$ and $C_{V}(T)$ exhibit qualitatively the same
behavior as the free fermion gas at the SM-SC QCP. Additionally,
because the residue $Z_f$ always takes a finite value, the fermion
spectral function should have a sharp peak. This feature can be
readily detected by ARPES experiments \cite{Damascelli03}.

\section{Summary and discussion \label{Sec:Summary}}

In summary, we have studied the influence of quantum critical
fluctuation of SC order parameter on the low-energy behavior of
fermions in 3D AWSM. Different from other quantum critical systems,
the anomalous dimension of anisotropic Weyl fermions flows to zero
quickly at low energies at the SM-SC QCP. As a consequence, the
fermion residue is always finite, indicating the validity of FL
description and the irrelevance of SC order parameter fluctuation.
It turns out that the crucial physics of the SM-SC quantum phase
transition can be captured by the simple mean-field analysis.

In a recent work, Yang \emph{et al.} \cite{Yang14A} demonstrated
that the long-range Coulomb interaction is an irrelevant
perturbation in 3D AWSM. Combining their results with ours, we
conclude that 3D AWSM is an unusual system in which the fermions are
extremely robust against repulsive long-range interactions. The
stability of the system is guaranteed by the special dispersion of
anisotropic Weyl fermions.

The 3D AWSM state could be realized either at the QCP between
band-insulator and ordinary 3D WSM, or at the QCP between
band-insulator and 3D topological insulator in some
non-centrosymmetric systems \cite{Yang14B}. For instance, it is
predicted that the 3D AWSM state may be achieved by applying
pressure to the compound BiTeI, in which the inversion symmetry is
broken \cite{Yang13, Bahramy12}. Experiments performed in pressured
BiTeI by means of X-ray powder diffraction and infrared spectroscopy
are consistent with these theoretical predictions \cite{Xi13}.
Recent Shubnikov-de Haas quantum oscillation experiments have
revealed evidence of a pressure-induced topological quantum phase
transition in BiTeI \cite{Park15}. Once superconductivity is induced
by certain mechanism in the mother 3D AWSM state, it should be
possible to measure some observable quantities, such as the fermion
spectral function and specific heat, to verify whether the
anisotropic Weyl fermions behave as free fermion gas at the SM-SC
QCP and also in the SM phase.

\section*{ACKNOWLEDGEMENTS}

We would like to acknowledge the financial support by the National
Natural Science Foundation of China under Grants 11574285 and
11504379. J.R.W. is partly supported by the Natural Science
Foundation of Anhui Province under Grant 1608085MA19. X.L. and
G.-Z.L. are partly supported by the Fundamental Research Funds for
the Central Universities (P. R. China) under Grant WK2030040085.

\appendix

\section{Superconducting quantum phase transition \label{Appendix:GapEquation}}

The system considered here is described by the partition function
\begin{eqnarray}
Z = \int D[\psi^{\dag},\psi]\exp\left(-\int_{0}^{\beta}d\tau\int
d^{3}\mathbf{x}\mathcal{L}\left[\psi^{\dag},\psi\right]\right),
\end{eqnarray}
where $\beta=1/k_{B}T$ and the Lagrangian is given by
\begin{eqnarray}
\mathcal{L} = \int\frac{d^{3}\mathbf{k}}{(2\pi)^{3}}
\psi_{\mathbf{k}}^{\dag}\partial_{\tau}\psi_{\mathbf{k}}+H,
\end{eqnarray}
where the sum over momentum is replaced by an integral.

We now define a four-component Nambu spinor
$\Psi=(\psi_{\mathbf{k}},\psi_{-\mathbf{k}}^{\dag})^{T}$. The
Fourier transformation in the imaginary-time space is
\begin{eqnarray}
\psi(\tau) = \frac{1}{\sqrt{\beta}}
\sum_{\omega_{n}}\psi(\omega_{n})e^{-i\omega_{n}\tau}.
\end{eqnarray}
Using the SC order parameter $\Delta_{s} = g\sum_{\mathbf{k}}\langle
\psi_{\mathbf{k}}(i\sigma_{2})\psi_{\mathbf{-k}}\rangle$, we
re-write the partition function in the form
\begin{eqnarray}
Z &=& \int D[\Psi^{\dag},\Psi,\Delta^{*},\Delta]\nonumber \\
&&\times\exp\left(-\int_{0}^{\beta}d\tau\int d^{3}\mathbf{x}
\mathcal{L}\left[\Psi^{\dag},\Psi,\Delta^{*},\Delta\right]\right),
\end{eqnarray}
where
\begin{eqnarray}
\mathcal{L}=\frac{1}{\beta}\sum_{\omega_{n}}\int \frac{d^{3}
\mathbf{k}}{(2\pi)^{3}} \psi_{\omega_{n},\mathbf{k}}^{\dag}
\mathcal{G}_{\omega_{n},\mathbf{k}}
\psi_{\omega_{n},\mathbf{k}}+\frac{|\Delta_{s}|^{2}}{2g}.
\end{eqnarray}
In the above expression, we have defined
\begin{eqnarray}
\mathcal{G}_{\omega_{n},\mathbf{k}}= \left(
\begin{array}{cccc}
\begin{smallmatrix}
-i\omega_{n}+Ak_{z}^{2}& vk_{+} & 0 &\Delta_{s} \\ vk_{-} &
-i\omega_{n}-Ak_{z}^{2} & -\Delta_{s} &0
\\ 0 & -\Delta_{s}^{*} & -i\omega_{n}+Ak_{z}^{2} & vk_{-} \\
\Delta_{s}^{*} & 0 & vk_{+} & -i\omega_{n}-Ak_{z}^{2}
\end{smallmatrix}
\end{array}\right),
\nonumber \\
\end{eqnarray}
where $k_{+}=k_{x}+ik_{y}$ and $k_{-}=k_{x}-ik_{y}$. To make a
mean-field analysis, we integrate out the fermionic degree of
freedom and then get an effective Lagrangian that contains only the
order parameter:
\begin{eqnarray}
\mathcal{L} &=& -T\sum_{\omega_{n}}\int\frac{d^{3}\mathbf{k}}{(2\pi)^3}
\textrm{ln}\;\textrm{det}\mathcal{G}_{\omega_{n},\mathbf{k}} +
\frac{|\Delta_{s}|^2}{2g}\nonumber
\\
&=&-T\sum_{\omega_{n}}\int\frac{d^{3}\mathbf{k}}{(2\pi)^3}
\textrm{ln}\left[\omega_{n}^{2}+E_{f}^2(\mathbf{k})+
|\Delta_{s}|^{2}\right]^{2} + \frac{|\Delta_{s}|^{2}}{2g}.\nonumber\\
\end{eqnarray}
Varying the action with respect to $|\Delta_{s}|$ yields the gap
equation represented by Eq.(3).

\section{Renormalization group calculations \label{Appendix:RGAnalysis}}

The mean-field calculation reveals a critical attraction strength
$g=g_c$. At this QCP, the gapless fermions and the gapless quantum
fluctuation of SC order parameter couple to each other. To examine
whether or not such an interaction have significant impact on the
quantum critical behavior, we now perform a detailed RG analysis.

At the SM-SC QCP, the partition function is
\begin{eqnarray}
Z = \int{D\phi}{D\phi^*}{D\psi}{D\psi^{\dag}}e^{-S},
\end{eqnarray}
where the action $S$ is given in the main text. Separating all the
field operators into slow and fast modes yields
\begin{eqnarray}
Z &=& \int{D\phi_{<}}{D\phi^*_{<}}{D\psi_{<}}{D\psi^{\dag}_{<}}
e^{-S_{0}^{<}} \nonumber \\
&& \times \int{D\phi_{>}}{D\phi^*_{>}}{D\psi_{>}}{D\psi^{\dag}_{>}}
e^{-S_{0}^{>}}e^{-S_{\phi^4}-S_{\phi\psi}} \nonumber \\
&=&Z_{0}^{>}\int{D\phi_{<}}{D\phi^*_{<}}{D\psi_{<}}{D\psi^{\dag}_{<}}
e^{-S_{0}^{<}}\langle{e^{-S_{\phi^4}-S_{\phi\psi}}}\rangle_{>},
\nonumber \\
\end{eqnarray}
where $\phi_{<}$ and $\psi_{<}$ are both slow modes and $\phi_{>}$
and $\psi_{>}$ are both fast modes. For simplicity, we have used the
following notations:
\begin{eqnarray}
Z_{0}^{>} &=& \int{D\phi_{>}}{D\phi^*_{>}}{D\psi_{>}}{D
\psi^{\dag}_{>}}e^{-S_{0}^{>}}, \\
\langle{e^{-S_{I}}}\rangle_{>} &=& \frac{1}{Z^{>}_{0}}
\int{D\phi_{>}}{D\phi^*_{>}}{D\psi_{>}}{D\psi^{\dag}_{>}}
e^{-S_{0}^{>}}e^{-S_{I}}.\nonumber \\
\end{eqnarray}
Here, $S_{I} = S_{\phi^4} + S_{\phi\psi}$. One can compute
$\langle{e^{-S_{I}}}\rangle_{>}$ by means of cumulant expansion
method \cite{Shankar94}. Up to the order of $O(h^4,\lambda^2)$, we
find that
\begin{eqnarray}
\langle e^{-S_{I}}\rangle_{>} = e^{-\left[\langle{S_{I}}\rangle_{>}
- \frac{1}{2} \langle{S_{I}^2}\rangle_{>} +
\frac{1}{6}\langle{S_{I}^3}\rangle_{>} -
\frac{1}{24}\langle{S_{I}^4}\rangle_{>}\right]}.
\end{eqnarray}

\textbf{Fermion self-energy corrections}

We will first consider the fermion self-energy corrections. For this
purpose, we compute $\Delta{S}_\psi = -\frac{1}{2}
\langle{S^{2}_{\psi\phi}}\rangle_{>}$, and find that
\begin{widetext}
\begin{eqnarray}
\Delta{S}_\psi&=&-\frac{h^2}{2}\int^{b\Lambda}d^{4}k{d^4q}
\int_{b\Lambda}^{\Lambda}d^{4}k'{d^4q'}
\langle(\phi^{*}\psi^{T}i\sigma_{2}\psi+H.c.)
(\phi'^{*}\psi'^{T}i\sigma_{2}\psi'+H.c.)\rangle_{>} \nonumber
\\&=& 4h^2\int^{b\Lambda}\frac{d\omega}{2\pi}\frac{d^3\mathbf{k}}{(2\pi)^3}
\psi^\dag\psi\int_{b\Lambda}^\Lambda\frac{{d\omega'}}{2\pi}\frac{d^3\mathbf{k}'}{(2
\pi)^3}[G_\phi(\omega+\omega',\mathbf{k}+\mathbf{k}')\sigma_2
G^T_\psi(\omega,\mathbf{k})\sigma_2],
\label{Eq:FermionSelfEnergyDef}
\end{eqnarray}
where for simplicity, we have defined
\begin{eqnarray}
\int^{b\Lambda}d^{4}k{d^4q}&=&\int^{b\Lambda}\prod_{i=1}^{2}
\frac{d\omega_{i}}{2\pi}\frac{d^{3}\mathbf{k_{i}}}{(2\pi)^{3}}
\frac{d\Omega}{2\pi} \frac{d^{3}\mathbf{q}}{(2\pi)^{3}}
\delta(\omega_{1}+\omega_{2}-\Omega) \delta^{3}(\mathbf{k_{1}} +
\mathbf{k_{2}} - \mathbf{q}),\nonumber \\
\int_{b\Lambda}^{\Lambda}d^{4}k'{d^4q'} &=& \int_{b\Lambda}^{\Lambda}\prod_{i=1}^{2}
\frac{d\omega'_{i}}{2\pi}\frac{d^{3} \mathbf{k_{i}}'}{(2\pi)^{3}}
\frac{d\Omega'}{2\pi}\frac{d^{3}\mathbf{q}'}{(2\pi)^{3}}
\delta(\omega'_{1}+\omega'_{2}-\Omega')
\delta^{3}(\mathbf{k_{1}}'+\mathbf{k_{2}}'-\mathbf{q}'),
\end{eqnarray}
here $\Lambda$ is the upper cutoff and $b = e^{-\ell}$.

We expand $\Delta{S}_\psi$ in powers of small external energy and
momentum, and then integrate over energy, which leads to
$\Delta{S}_\psi = \Delta{S}_\psi^1 + \Delta{S}_\psi^2 +
\Delta{S}_\psi^3$, where
\begin{eqnarray}
\Delta{S}_\psi^1 &=& \int^{b\Lambda}\frac{d\omega}{2\pi}
\frac{d^3\mathbf{k}}{(2\pi)^3}(-i\omega\sigma_{0}) \psi^\dag\psi
\int_{b\Lambda}^{\Lambda}{d}k_{\perp}'d|k_{z}'|k_{\perp}'
\frac{2h^2F_{1}}{(2\pi)^2}, \nonumber \\
\Delta{S}_\psi^2 &=& \int^{b\Lambda}\frac{d\omega}{2\pi}
\frac{d^3\mathbf{k}}{(2\pi)^3} c_{f}\sigma_{1}k_{x} \psi^\dag\psi
\int_{b\Lambda}^{\Lambda}{d}k_{\perp}'d|k_{z}'|\frac{h^2
F_{2}}{(2\pi)^2} + \int^{b\Lambda}\frac{d\omega}{2\pi}
\frac{d^3\mathbf{k}}{(2\pi)^3}c_{f} \sigma_{2}k_{y}\psi^\dag
\psi\int_{b\Lambda}^{\Lambda}{d}k_{\perp}'d|k_{z}'|
\frac{h^2F_{2}}{(2\pi)^2}, \nonumber \\
\Delta{S}_\psi^3 &=& \int^{b\Lambda}\frac{d\omega}{2\pi}
\frac{d^3\mathbf{k}}{(2\pi)^3}(Ak^{2}_{z}\sigma_{3}) \psi^{\dag}\psi
\int_{b\Lambda}^{\Lambda}{d}k'_{\perp}d|k'_{z}|
\frac{h^2F_{3}}{(2\pi)^2} - \int^{b\Lambda}\frac{d\omega}{2\pi}
\frac{d^3\mathbf{k}}{(2\pi)^3}(Ak^{2}_{z}\sigma_{3})\psi^{\dag}\psi
\int_{b\Lambda}^{\Lambda}{d} k'_{\perp}d|k'_{z}|
\frac{h^2F_{4}}{(2\pi)^2}.\nonumber
\end{eqnarray}
\end{widetext}
Here, the four functions $F_{1,2,3,4}$ are given by
\begin{eqnarray}
F_{1}&=&\frac{1}{E_{b}(\mathbf{k}')[E_{b}(\mathbf{k}') +
E_{f}(\mathbf{k}')]^2}, \\
F_{2}&=&\frac{c^{2}_{\perp}k'^{3}_{\perp}[E_{f}(\mathbf{k}') +
2E_{b}(\mathbf{k}')]}{E_{f}(\mathbf{k}')E_{b}^3(\mathbf{k}')[E_{f}(\mathbf{k}')
+ E_{b}(\mathbf{k}')]^2}, \\
F_{3}&=&\frac{c_{bz}^2k'^{2}_{z}k'_{\perp}[E_{f}(\mathbf{k}') +
2E_{b}(\mathbf{k}')]}{E_{f}(\mathbf{k}')E^3_{b}(\mathbf{k}')[E_{b}(\mathbf{k}')
+ E_{f}(\mathbf{k}')]^2}, \\
F_{4}&=&\frac{c_{bz}^{4}k'^4_zk'_{\perp}[3E^2_{f}(\mathbf{k}') +
9E_{f}(\mathbf{k}')E_{b}(\mathbf{k}')+8E^2_{f}(\mathbf{k}')]}
{E_{f}(\mathbf{k}') E^5_{b}(\mathbf{k}')[E_{f}(\mathbf{k}') +
E_{b}(\mathbf{k}')]^3},\nonumber \\
\end{eqnarray}
where
\begin{eqnarray}
E_{f}(\mathbf{k}')&=&\sqrt{c_{f}^{2}k_{\perp}'^{2}+A^{2}k_{z}'^{4}}, \\
E_{b}(\mathbf{k}')&=&\sqrt{c_{b\perp}^{2}k'^{2}_{\perp} +
c_{bz}^{2}k_{z}'^{2}}.
\end{eqnarray}
A constant term that is independent of external energy and momenta
has been dropped during the calculation. To proceed, we find it
convenient to employ the following transformations
\begin{eqnarray}
E = \sqrt{c_{f}^{2}k_{\perp}'^{2} + A^{2}k_{z}'^{4}}, \quad \delta =
\frac{c_{f}k_{\perp}'}{Ak_{z}'^2}, \label{Eq:TransformationA}
\end{eqnarray}
which are equivalent to
\begin{eqnarray}
k_{\perp}' = \frac{E\delta}{c_{f}\sqrt{1+\delta^{2}}}, \quad
|k_{z}'| = \frac{\sqrt{E}}{\sqrt{A}(1+\delta^{2})^{1/4}}.
\label{Eq:TransformationB}
\end{eqnarray}
The integral measure satisfies the relation
\begin{eqnarray}
dk_{\perp}'d|k_{z}'| = \frac{\sqrt{E}}{2c_{f}
\sqrt{A}(1+\delta^{2})^{\frac{3}{4}}}dE d\delta.
\label{Eq:TransformationC}
\end{eqnarray}

After accomplishing the above transformations, the next step is to
integrate out all the fast modes, which gives rise to
\begin{widetext}
\begin{eqnarray}
\Delta{S}_\psi&=&\int^{b\Lambda}\frac{d\omega}{2\pi}
\frac{d^3\mathbf{k}}{(2\pi)^3}\psi^{\dag} \psi(-i\omega
\sigma_{0})C_{1}\ell + \int^{b\Lambda}\frac{d\omega}{2\pi}
\frac{d^3\mathbf{k}}{(2\pi)^3} \psi^{\dag}\psi
c_{f}\sigma_{1}k_{x}C_{2}\ell \nonumber \\
&&+\int^{b\Lambda}\frac{d\omega}{2\pi}\frac{d^3\mathbf{k}}{(2\pi)^3}
\psi^{\dag}\psi c_{f}\sigma_{2}k_{y}C_{2}\ell + \int^{b\Lambda}
\frac{d\omega}{2\pi}\frac{d^3\mathbf{k}}{(2\pi)^3}\psi^{\dag}\psi
(Ak'^{2}_{z}\sigma_{3})C_{3}\ell.
\end{eqnarray}
The three constants $C_{1,2,3}$ are given by
\begin{eqnarray}
C_{1}&=&\frac{h'^2}{(2\pi)^2}\int_{0}^{+\infty}
\frac{\eta_{A}^{-1}(1+\delta^2)^{1/4}
\delta{d\delta}}{\sqrt{F_{1}}F_{2}^2}, \\
C_{2}&=&\frac{h'^2}{(2\pi)^2} \int_{0}^{+\infty}
d\delta\frac{\varsigma\eta_{A}^{-1}\eta_{B}^2 \delta^3
(1+\delta^{2})^{1/4}}{2F_{1}^{3/2}F_{2}^2}+\frac{h'^2}{(2\pi)^2}
\int_{0}^{+\infty}d\delta \frac{\eta_{B}^2 \delta^3
\sqrt{\varsigma}}{(1+\delta^{2})^{1/4}F_{1}F_{2}^2}, \\
C_{3} &=& \frac{h'^2}{(2\pi)^2} \int_{0}^{+\infty} d\delta
\frac{\eta_{A}^{-1}\delta (1+\delta^{2})^{1/4}}{2F_{1}^{3/2}F_{2}^2}
+\frac{h'^2}{(2\pi)^2}\int_{0}^{+\infty}d\delta
\frac{\delta}{(1+\delta^{2})^{1/4}\sqrt{\varsigma}
F_{1}F_{2}^2} \nonumber \\
&&-\frac{h'^2}{(2\pi)^2}\int_{0}^{+\infty}d\delta
\frac{3\eta_{A}^{-1}\delta(1+\delta^{2})^{7/4}}{2F_{1}^{5/2}F_{2}^{3}}
-\frac{h'^2}{(2\pi)^2}\int_{0}^{+\infty}{d\delta}
\frac{9\varsigma^{-1/2}\delta(1+\delta^2)^{5/4}}{2F_{1}^{2}F_{2}^{3}}
\nonumber \\
&&-\frac{h'^2}{(2\pi)^2}\int_{0}^{+\infty}d\delta \frac{4\eta_{A}
\varsigma^{-1}\delta}{(1+\delta^{2})^{3/4} F_{1}^{3/2}F_{2}^{3}},
\end{eqnarray}
where
\begin{eqnarray}
F_{1}&=&\sqrt{1+\delta^2} + \varsigma\eta_{B}^2\delta^2, \\
F_{2}&=&\sqrt{1+\delta^2}+\frac{\eta_{A}}{\sqrt{\varsigma}}
\sqrt{\sqrt{1+\delta^2} +\varsigma\eta_{B}^2\delta^2}.
\end{eqnarray}
In the above calculation, we have defined three new parameters:
\begin{eqnarray}
\varsigma = \frac{A\Lambda}{c^2_{f}},\quad \eta_{A} =
\frac{c_{bz}}{c_{f}},\quad \eta_{B} = \frac{c_{b\perp}}{c_{bz}}.
\end{eqnarray}

\textbf{Boson self-energy corrections}

We then consider the corrections to the action of boson field. In
particular, we need to compute $\Delta{S}_\phi = -\frac{1}{2}
\langle{S^{2}_{\psi\phi}}\rangle_{>}$. It is straightforward to get
\begin{eqnarray}
\Delta{S}_\phi&=&-\frac{h^2}{2}\int^{b\Lambda}d^{4}k{d^4q}
\int_{b\Lambda}^{\Lambda}d^{4}k'{d^4q'} \langle(\phi^{*}
\psi^{T}i\sigma_{2}\psi+H.c.)(\phi'^{*}\psi'^{T}i\sigma_{2}\psi' +
H.c.)\rangle_{>} \nonumber \\
&=&2h^{2}\int^{b\Lambda}\frac{d\Omega}{2\pi}
\frac{d^{3}\mathbf{q}}{(2\pi)^{3}} \phi_{s}\phi_{s}^{*}
\int_{b\Lambda}^{\Lambda}\frac{d\omega'}{2\pi}
\frac{d^{3}\mathbf{k}'}{(2\pi)^{3}} \mathrm{Tr}\left[\sigma_{2}
G_{\psi}^{T}(\omega',\mathbf{k}')\sigma_{2}G_{\psi}(\Omega+\omega',
\mathbf{q}+\mathbf{k}')\right]. \label{Eq:FermionSelfEnergyDef}
\end{eqnarray}
Similarly, we obtain $\Delta{S}_\phi = \Delta{S}_\phi^1 +
\Delta{S}_\phi^2 + \Delta{S}_\phi^3$, where
\begin{eqnarray}
\Delta{S}_\phi^1 &=& \frac{h^2}{(2\pi)^2}\int^{b\Lambda}
\frac{d\Omega}{2\pi} \frac{d^{3}\mathbf{q}}{(2\pi)^{3}}
\phi_{s}\phi_{s}^{*} \Omega^2
\int_{b\Lambda}^{\Lambda}{d}k'_{\perp}d|k'_{z}| k'_{\perp}
\frac{1}{2E^3_{f}(\mathbf{k}')}, \\
\Delta{S}_\phi^2 &=& \frac{h^2}{(2\pi)^2}\int^{b\Lambda}
\frac{d\Omega}{2\pi}\frac{d^{3}\mathbf{q}}{(2\pi)^{3}}
\phi_{s}\phi_{s}^{*} c_{b\perp}^2 q_{\perp}^2
\int_{b\Lambda}^{\Lambda}{d}k'_{\perp}d|k'_{z}| k'_{\perp}
\frac{c_{f}^2}{c_{b\perp}^2 E^3_{f}(\mathbf{k}')}
\nonumber\\
&&-\frac{h^2}{(2\pi)^2}\int^{b\Lambda}\frac{d\Omega}{2\pi}
\frac{d^{3}\mathbf{q}}{(2\pi)^{3}} \phi_{s}\phi_{s}^{*} c_{b\perp}^2
q_{\perp}^2 \int_{b\Lambda}^{\Lambda}{d} k'_{\perp} d|k'_{z}|
k'_{\perp} \frac{3c_{f}^{4}{k}_{\perp}'^2}{4c_{b\perp}^2
E^5_{f}(\mathbf{k}')}, \\
\Delta{S}_\phi^3 &=& \frac{h^2}{(2\pi)^2}
\int^{b\Lambda}\frac{d\Omega}{2\pi}
\frac{d^{3}\mathbf{q}}{(2\pi)^{3}} \phi_{s}\phi_{s}^{*}
c_{bz}^2q_{z}^2\int_{b\Lambda}^{\Lambda}{d}k'_{\perp}d|k'_{z}|
k'_{\perp} \frac{5A^2 k'^2_{z}}{c_{bz}^2 E^3_{f}(\mathbf{k}')}
\nonumber\\
&&-\frac{h^2}{(2\pi)^2}\int^{b\Lambda}\frac{d\Omega}{2\pi}
\frac{d^{3} \mathbf{q}}{(2\pi)^{3}} \phi_{s}\phi_{s}^{*} c_{bz}^2
q_{z}^2 \int_{b\Lambda}^{\Lambda}{d}k'_{\perp}d|k'_{z}| k'_{\perp}
\frac{6{A}^{4} k_{z}'^6}{c_{bz}^2 E^5_{f}(\mathbf{k}')}.
\end{eqnarray}
Similarly, we now make the transformations given by
Eqs.~(\ref{Eq:TransformationA})- (\ref{Eq:TransformationC}), and
then integrate over $E$ in the range of $b\Lambda<E<\Lambda$ and
integrate over $\delta$ in the range of $0<\delta<\infty$. After
performing such calculations, we obtain
\begin{eqnarray}
\Delta{S}_\phi = \int^{b\Lambda}\frac{d\Omega}{2\pi}\frac{d^{3}q}{(2\pi)^{3}}
\phi_{s}\phi_{s}^{*}\Omega^2C_{4}\ell + \int^{b\Lambda}\frac{d\Omega}{2\pi}
\frac{d^{3}q}{(2\pi)^{3}}\phi_{s}\phi_{s}^{*} c_{b\perp}^2
q_{\perp}^2C_{5}\ell + \int^{b\Lambda} \frac{d\Omega}{2\pi}
\frac{d^{3}q}{(2\pi)^{3}}\phi_{s} \phi_{s}^{*} c_{bz}^2
q_{z}^2 C_{6}\ell.
\end{eqnarray}
\end{widetext}
In this expression, we have defined three constants:
\begin{eqnarray}
C_{4}&=&\frac{h'^{2}}{(2\pi)^{2}\sqrt{\varsigma}},\\
\quad
C_{5}&=&\frac{4h'^{2}}{5(2\pi)^{2}\eta_{B}^2\sqrt{\varsigma}},\\
\quad
C_{6}&=&\frac{34\sqrt{\varsigma}h'^{2}}{21(2\pi)^{2}\eta_{A}^2}.
\end{eqnarray}

\textbf{Renomarlization of $\lambda$ at $O(\lambda^2)$}\\

At the order of $O(\lambda^2)$, $\Delta{S}_{\phi^4}^1=
-\frac{1}{2}\langle{S^2_{\phi}}\rangle_{>}$ is given by
\begin{eqnarray}
\Delta{S}_{\phi^4}^1 &=& -\frac{5}{2}\lambda^{2} \int^{b\Lambda} \prod_{i=1}^{4}
\frac{d\Omega_{i}}{2\pi}\frac{d^{3}\mathbf{q_{i}}}{(2\pi)^{3}}
\Delta(\Omega)\Delta(\mathbf{q})|\phi_{s}|^4 \nonumber \\
&&\times\int_{b\Lambda}^{\Lambda}\frac{d\Omega'}{2\pi}\frac{d^{3}\mathbf{q}'}{(2\pi)^{3}}
G_{\phi}(\Omega',\mathbf{q}')G_{\phi}(\Omega',\mathbf{q}').
\end{eqnarray}
Integrating over $\Omega'$, we find that
\begin{eqnarray}
\Delta{S}_{\phi^4}^1 &=& -\frac{5\lambda^{2}}{8(2\pi)^2}
\int^{b\Lambda}\prod_{i=1}^{4}\frac{d\Omega_{i}}{2\pi} \frac{d^{3}
\mathbf{q_{i}}}{(2\pi)^{3}}\Delta(\Omega)\Delta(\mathbf{q})
|\phi_{s}|^4 \nonumber \\
&&\times \int^{\Lambda}_{b\Lambda}{q'_{\perp}}dq'_{\perp}dq'_{z}
\frac{1}{E_{b}(\mathbf{q}')^3}.
\end{eqnarray}
Using the following transformations
\begin{eqnarray}
E = \sqrt{c_{b\perp}^{2}q_{\perp}'^{2}+c_{bz}^{2}q_{z}'^{2}}, \quad
\delta = \frac{c_{b\perp}q_{\perp}'}{c_{bz}|q_{z}'|},
\end{eqnarray}
where $\delta\in(0,+\infty)$, it is easy to get
\begin{eqnarray}
q_{\perp}' &=& \frac{E\delta}{c_{b\perp}\sqrt{1+\delta^{2}}},\\
\quad |q_{z}'| &=& \frac{E}{c_{bz}\sqrt{1+\delta^{2}}},\\
\quad dq' d|q_{z}'| &=& \frac{E}{c_{b\perp} c_{bz}(1 + \delta^{2})}
dE d\delta.
\end{eqnarray}
Finally we obtain
\begin{eqnarray}
\Delta{S}_{\phi^4}^1 = \int^{b\Lambda}\prod_{i=1}^{4}
\frac{d\Omega_{i}}{2\pi}\frac{d^{3}\mathbf{q_{i}}}{(2\pi)^{3}}
\Delta(\Omega)\Delta(\mathbf{q}) \frac{C_{7}}{4} \lambda\ell
|\phi_{s}|^4,
\end{eqnarray}
where
\begin{eqnarray}
C_{7} = -\frac{5\lambda'}{2(2\pi)^2 \eta_{A} \eta_{C}^2},
\end{eqnarray}
with
\begin{eqnarray}
\eta_{C}=\frac{c_{b\perp}}{c_{f}}.
\end{eqnarray}

\textbf{Renormalization of $\lambda$ at $O(h^4)$}\\

At the order of $O(h^4)$, $\Delta{S}_{\phi^4}^2 = -\frac{1}{24}
\langle{S^4_{\phi\psi}}\rangle_{>}$ is
\begin{eqnarray}
\Delta{S}_{\phi^4}^2 &=&
4h^{4}\int^{b\Lambda}d^4q |\phi_s|^4 \int^{\Lambda}_{b\Lambda}
\frac{d\omega'}{2\pi}\frac{d^{3}\mathbf{k}'}{(2\pi)^{3}}
\nonumber \\
&&\times \mathrm{Tr}[(i\sigma_{2})G^{T}(\omega',\mathbf{k}')
(-i\sigma_{2})G(-\omega',-\mathbf{k}') \nonumber\\
&& \times i\sigma_{2}G^{T}(\omega',\mathbf{k}')(-i\sigma_{2})
G(-\omega',-\mathbf{k}')],
\end{eqnarray}
where for simplicity we set
\begin{eqnarray}
\int^{b\Lambda}d^4q = \int^{b\Lambda}\prod_{i=1}^{4}\frac{d\Omega_{i}}{2\pi}
\frac{d^{3}\mathbf{q_{i}}}{(2\pi)^{3}}
\Delta(\Omega)\Delta(\mathbf{q}).
\end{eqnarray}

After carrying out integrations, we have
\begin{eqnarray}
\Delta{S}_{\phi^4}^2 =
\int^{b\Lambda}d^4q|\phi_{s}|^4\frac{C_{8}}{4}\lambda\ell,
\end{eqnarray}
where
\begin{eqnarray}
C_{8} = \frac{8h'^{4}}{(2\pi)^{2}\lambda'\sqrt{\varsigma}}.
\end{eqnarray}
Thus the total correction to $\lambda$ is given by
\begin{eqnarray}
\Delta\lambda = \left(C_{7}+C_{8}\right)\lambda\ell.
\end{eqnarray}

\section{Derivation of the RG equations}

Adding fermion self-energy to the free action yields
\begin{eqnarray}
S_{\psi}' &=& \int\frac{d\omega}{2\pi}
\frac{d^{3}k}{(2\pi)^{3}}{\psi}^{\dagger} \psi\mathcal{L}_{\psi} +
\Delta{S}_\psi, \\
\mathcal{L}_\psi &=& -i\omega\sigma_{0}+c_{f}
\left(k_{x}\sigma_{1}+k_{y}\sigma_{2}\right) + Ak^{2}_{z}\sigma_{3}.
\end{eqnarray}
Using the following scale transformations
\begin{eqnarray}
k_{x}&=&e^{-\ell}k_{x}', \label{Eq:Scalingkx}
\\
k_{y}&=&e^{-\ell}k_{y}', \label{Eq:Scalingky} \\
k_{z}&=&e^{-\frac{\ell}{2}}k_{z}', \label{Eq:Scalingkz} \\
\omega&=&e^{-\ell}\omega', \label{Eq:Scalingomega} \\
\psi&=&e^{(\frac{9}{4}-\frac{C_{1}}{2})\ell}\psi',
\label{Eq:Scalingpsi} \\
c_{f}&=&e^{(C_{1}-C_{2})\ell}c_{f}', \label{Eq:Scalingcf} \\
A&=&e^{(C_{1}-C_{3})\ell}A', \label{Eq:ScalingA}
\end{eqnarray}
we re-write the action in the form
\begin{eqnarray}
S_{\psi'}'&=&\int\frac{d\omega'}{2\pi} \frac{d^{3}
\mathbf{k}'}{(2\pi)^{3}}{\psi}'^{\dagger}\psi'\mathcal{L}'_{\psi'}, \\
\mathcal{L}'_{\psi'} &=& -i\omega'\sigma_{0}+c_{f}'(k_{x}'\sigma_{1} +
k_{y}'\sigma_{2})+A'k'^{2}_{z}\sigma_{3}.
\end{eqnarray}

For the boson field, the renormalized action is given by
\begin{eqnarray}
S_{\phi}' &=& \frac{1}{2}\int\frac{d\Omega}{2\pi}
\frac{d^{3}\mathbf{q}}{(2\pi)^{3}}\phi^*\phi\mathcal{L}_\phi +
\Delta{S}_\phi,
\label{Eq:ActionBosonOneLoopCorrect} \\
\mathcal{L}_\phi &=& \Omega^{2} + c_{b\perp}^{2}
\mathbf{q}^{2}_{\perp} + c_{z}^{2}q_{z}^{2}.
\end{eqnarray}
Employing the transformations
(\ref{Eq:Scalingkx})-(\ref{Eq:Scalingomega}), along with
\begin{eqnarray}
\phi &=& e^{(\frac{11}{4}-\frac{C_{4}}{2})\ell}\phi',
\label{Eq:Scalingphi} \\
c_{b\perp}&=&e^{(\frac{C_{4}}{2} -
\frac{C_{5}}{2})\ell}c_{b\perp}', \label{Eq:ScalingcbBot}
\\
c_{bz}&=&e^{(\frac{C_{4}}{2}-\frac{C_{6}}{2} -
\frac{1}{2})\ell}c_{bz}', \label{Eq:Scalingcbc}
\end{eqnarray}
we can re-write the above action as
\begin{eqnarray}
S_{\phi'}' &=& \frac{1}{2}\int \frac{d\Omega'}{2\pi}\frac{d^{3}
\mathbf{q}'}{(2\pi)^{3}}\phi'^{*}\phi'\mathcal{L}'_{\phi'},\\
\mathcal{L}'_{\phi'} &=& \Omega'^{2}+c_{b\perp}'^{2}
\mathbf{q}'^{2}_{\perp} + c_{bz}'^{2}q_{z}'^{2}.
\end{eqnarray}

Including one-loop corrections to the action of four-boson coupling
leads to
\begin{eqnarray}
S_{\phi^{4}}'&=&\frac{\lambda+\Delta\lambda}{4}\int \prod_{i=1}^{4}
\frac{d\Omega_{i}}{2\pi}\frac{d^{3}\mathbf{q}_{i}}{(2\pi)^{3}}
\Delta(\Omega)\Delta(\mathbf{q})|\phi|^{4}\nonumber \\
&\approx&\frac{\lambda e^{(C_{7}+C_{8})\ell}}{4}\int\prod_{i=1}^{4}
\frac{d\Omega_{i}}{2\pi}\frac{d^{3}\mathbf{q}_{i}}{(2\pi)^{3}}
\Delta(\Omega)\Delta(\mathbf{q})|\phi|^{4}.\nonumber \\
\label{Eq:ActionFourBosonOneLoopCorrect}
\end{eqnarray}
We use the transformations
Eqs.~(\ref{Eq:Scalingkx})-(\ref{Eq:Scalingomega}),
Eq.~(\ref{Eq:Scalingphi}), and the extra transformation
\begin{eqnarray}
\lambda = \lambda'e^{\left(2C_{4} - \frac{1}{2} -
C_{7}-C_{8}\right)\ell}, \label{Eq:Scalinglambda}
\end{eqnarray}
and then find that
\begin{eqnarray}
S_{\phi'^{4}}' &=& \frac{\lambda'}{4}\int\prod_{i=1}^{4}
\frac{d\Omega_{i}'}{2\pi}\frac{d^{3}\mathbf{q}_{i}'}{(2\pi)^{3}}
\Delta(\Omega')\Delta(\mathbf{q}')|\phi'|^{4}.
\end{eqnarray}

The Yukawa-coupling can be treated by employing the same
calculational steps. In particular, we invoke the transformations
Eqs.~(\ref{Eq:Scalingkx})-(\ref{Eq:Scalingpsi}),
Eq.~(\ref{Eq:Scalingphi}), and an additional transformation
\begin{eqnarray}
h = h'e^{\left(\frac{C_{4}}{2}+C_{1}-\frac{1}{4}\right)\ell}.
\label{Eq:Scalingh}
\end{eqnarray}
After straightforward calculations, we finally obtain the following
action for the Yukawa-coupling
\begin{eqnarray}
S_{\psi'\phi'} &=& h'\int\prod_{i=1}^{2}\frac{d\omega'_{i}}{2\pi}
\frac{d^{3} \mathbf{k}'_{i}}{(2\pi)^{3}}\frac{d\Omega'}{2\pi}
\frac{d^{3}\mathbf{q}'}{(2\pi)^{3}}\nonumber \\
&& \times \delta(\omega_{1}'-\omega_{2}'+\Omega')
\delta^{3}(\mathbf{k}_{1}'-\mathbf{k}_{2}'+ \mathbf{q}') \nonumber
\\
&&\times (\phi'^{*}\psi'^{T}i\sigma_{2}\psi' + \mathrm{H.c.}).
\end{eqnarray}

By employing the transformations Eqs.~(\ref{Eq:Scalingcf}),
(\ref{Eq:ScalingA}), (\ref{Eq:ScalingcbBot}), (\ref{Eq:Scalingcbc}),
(\ref{Eq:Scalinglambda}), and (\ref{Eq:Scalingh}), we have derived
the coupling RG equations Eqs.~(\ref{Eq:VRGcf})-(\ref{Eq:VRGh})
presented in the main text of the paper.

\end{document}